\documentclass[final,twocolumn]{IEEEtran}
\usepackage{mathrsfs} \usepackage{graphicx,color} \usepackage{bbm} \usepackage{multirow}
\usepackage{lipsum}

\usepackage{amsmath} 
\usepackage{amssymb}
\usepackage{amsfonts}
\usepackage{latexsym}
\usepackage{amsthm}
\usepackage{varioref}
\usepackage{listings}
\usepackage{float}
\usepackage{rotate}
\usepackage{epsfig}
\usepackage{psfrag}
\usepackage{mathtools}
\usepackage{longtable}

\usepackage{graphicx, color}

\definecolor{darkred}{rgb}{0.9,0,0.3}
\definecolor{darkblue}{rgb}{0,0.3,0.9}

   \let\G=\Gamma

\def\ie{{i.e. }}\def\eg{{e.g. }}

\def\Re{{\rm Re}\,}\def\Im{{\rm Im}\,}

 \def\HHH{{\bf H}} \def\AAA{{\bf A}} 
\def\BBB{{\bf B}}

\numberwithin{equation}{section}

\renewcommand{\b}[1]{\boldsymbol{\mathrm{#1}}} 
\renewcommand{\cal}{\mathcal}

\newcommand \M {\textbf{M}}
\newcommand \C {\textbf{C}}
\newcommand \X {\textbf{X}}
\newcommand \R {\textbf{R}}

\newcommand \B {\textbf{B}}
\renewcommand \O {\textbf{O}}

\newcommand \Tr {\text{Tr}}
\newcommand \In {\b I_{N}}
\newcommand \It {\b {I}_{T}}
\newcommand{\bxp}{\boxplus}
\newcommand{\bxt}{\boxtimes}
\newcommand{\wh}{\widehat}
\newcommand{\argmin}{\operatornamewithlimits{argmin}}

\renewcommand{\G}{\b G}
\newcommand{\stj}{\mathfrak{g}}
\newcommand{\hil}{\mathfrak{h}}
\newcommand{\btr}{\cal{B}}
\newcommand{\rtr}{\cal{R}}
\newcommand{\wtr}{\cal{W}}
\newcommand{\ttr}{\cal{T}}
\newcommand{\str}{\cal{S}}

\newcommand{\ii}{\mathrm{i}}
\newcommand{\dd}{\mathrm{d}}

\newcommand*{\deq}{\mathrel{\vcenter{\baselineskip0.65ex \lineskiplimit0pt \hbox{.}\hbox{.}}}=}

\renewcommand{\le}{\leqslant}

\renewcommand{\ge}{\geqslant}
\renewcommand{\epsilon}{\varepsilon}

\newcommand{\pB}[1]{\Bigl({#1}\Bigr)}

\newcommand{\pBB}[1]{\Biggl({#1}\Biggr)}

\newcommand{\qb}[1]{\bigl[{#1}\bigr]}

\newcommand{\qBB}[1]{\Biggl[{#1}\Biggr]}

\newcommand{\normb}[1]{\bigl\lVert #1 \bigr\rVert}

\newcommand{\avg}[1]{\langle #1 \rangle}

\newcommand{\avgBB}[1]{\Biggl\langle #1 \Biggr\rangle}

\newcommand{\bra}[1]{\left\langle {#1} \right\lvert}
\newcommand{\ket}[1]{\left| {#1} \right\rangle}
\newcommand{\braket}[2]{\langle{#1} \lvert {#2}\rangle}

\newcommand{\innerb}[3]{\bigl\langle{#1} \bigl\lvert {#2} \bigl\lvert {#3} \bigl\rangle\,}

\begin{document}

\title{Rotational invariant estimator for general noisy matrices}


\author{\IEEEauthorblockN{Jo{\"e}l Bun\IEEEauthorrefmark{1} \IEEEauthorrefmark{2} \IEEEauthorrefmark{3},
Romain Allez\IEEEauthorrefmark{4}, Jean-Philippe Bouchaud \IEEEauthorrefmark{1} and
Marc Potters \IEEEauthorrefmark{1}} \\
\IEEEauthorblockA{\IEEEauthorrefmark{1}Capital Fund Management, 23 rue de l'Universit\'e, 75\,007 Paris, France,\\
\IEEEauthorrefmark{2}LPTMS, CNRS, Univ. Paris-Sud, Universit{\'e} Paris-Saclay, 91405 Orsay, France,\\
\IEEEauthorrefmark{3}Leonard de Vinci P{\^o}le Universitaire, Finance Lab, 92916 Paris La D{\'e}fense, France,\\
\IEEEauthorrefmark{4}Weierstrass Institute, Mohrenstr. 39, 10117 Berlin, Germany}}

\maketitle




\begin{abstract} 
We investigate the problem of estimating a given real symmetric signal matrix $\C$ from a noisy observation matrix $\M$ in the limit of large dimension. We consider the case where the noisy measurement $\M$ comes either from an arbitrary additive or multiplicative rotational invariant perturbation. We establish, using the Replica method, the asymptotic global law estimate for three general classes of noisy matrices, significantly extending previously obtained results. We give exact results concerning the asymptotic deviations (called {\it overlaps}) of the perturbed eigenvectors away from the true ones, and we explain how to use these overlaps to ``clean'' the noisy eigenvalues of $\M$. We provide some numerical checks for the different estimators proposed in this paper and we also make the connexion with some well known results of Bayesian statistics. 
\end{abstract}




\section{Introduction}

One of the most challenging problem in modern statistical analysis is to extract a true signal from noisy observations in data sets of very large dimensionality. Be it in physics, 
genomics, engineering or finance, scientists are confronted with datasets where the sample size $T$ and the number of variables $N$ are both very large, but with an observation ratio $q = N/T$ 
that is not small compared to unity. This setting is known in the literature as the high-dimensional limit and differs from the traditional large $T$, fixed $N$ situation (i.e. $q \to 0$), meaning that classical results of multivariate statistics do not necessarily apply. 

However, when one deals with very large random matrices (such as covariance matrices), one expects the spectral measure of the matrix under scrutiny to exhibit some universal properties, which are independent of the specific realization of the matrix itself. This property is at the core of Random Matrix Theory (RMT), which provides a very precise description of the convergence of the spectral measure for a very large class of random matrices. Perhaps the two most influential results are Wigner's semicircle law \cite{wigner1951statistical} and Mar{\v c}enko and Pastur's theorem \cite{marchenko1967distribution}. As far as inference is concerned, the latter result is arguably the cornerstone result of RMT in the sense that it gives theoretical tools to understand why classical estimators are insufficient and is now at the heart of many applications in this field (for reviews, see e.g. \cite{tulino2004random,johnstone2009statistical,bouchaud2009financial,couillet2011random} or more recently \cite{paul2014random} and references therein). 

In this paper, we consider the statistical problem of a $N \times N$ matrix $\C$ which stands for the unknown signal that one would like to estimate from the noisy measurement of a $N \times N$ matrix $\M$ in the limit of large dimension $N \rightarrow \infty$. A natural question in statistics is to find an estimator $\widehat{\b\Xi}(\M)$ of the true signal $\C$ that depends on the dataset $\M$ we have.
The true matrix $\C$ is unknown and we do not have any particular insights on its components (the eigenvectors). 
Therefore we would like our estimator $\widehat{\b\Xi}(\M)$ to be constructed
in a rotationally invariant way from the noisy observation $\M$ that we have.  
In simple terms, this means that there is no 
privileged direction in the $N$-dimensional space that would allow one to bias the eigenvectors of the estimator 
$\widehat{\b\Xi}(\M)$ in some special directions. More formally, the 
estimator construction must obey: 
\begin{equation}
\label{RIH}
\b\Omega \, \widehat{\b\Xi}(\M) \,\b\Omega^{\dag} = \widehat{\b\Xi}(\b\Omega \, \M\,\b\Omega^{\dag}),
\end{equation}
for any rotation matrix $\b\Omega$. Any estimator satisfying Eq. (\ref{RIH}) will be referred to as a Rotational Invariant Estimator (RIE).  In this case, it turns out that 
the eigenvectors of the estimator $\widehat{\b\Xi}(\M)$ have to be the same as those of the noisy matrix $\M$ \cite{stein1975estimation,takemura1983orthogonally}. 
As we will show in Section \ref{section2}, this implies that the best possible estimator 
$\widehat{\b\Xi}(\M)$ depends on the {\it overlaps} (i.e. the squared scalar product) between the eigenvectors of $\C$ and those of $\M$. These overlaps turn out to be fully computable in the
large $N$ limit, using tools from Random Matrix Theory, for a wide class of noise sources, much beyond the usual Gaussian models.

The study of the eigenvectors for statistical purposes is in fact quite a recent topic in random matrices. For 
sample covariance matrices, such considerations have been studied in \cite{bai2007asymptotics} and \cite{ledoit2011eigenvectors}. In the latter paper, the notion of overlap and optimal (oracle) estimator are treated in great details. As far as we know, this is the only paper in the literature where the oracle estimator is related to random matrices. Beside the sample covariance matrix, the problem of the overlap for a Gaussian matrix with an external source (also named as deformed Wigner ensemble) has been treated first in \cite{allez2013eigenvectors} and then reconsidered in a more general setting in \cite{allez2014eigenvectors} using Dyson Brownian motions (for an early -- but extremely  brief -- mention of these overlaps, see \cite[Section 6.2]{biane2003free}). However, no mention on how to clean the `noisy' matrix was given in \cite{allez2013eigenvectors,allez2014eigenvectors}. It is the gap we hope to fill here. We also extend these results to a much broader class of random perturbations, which, to the best of our knowledge, was not considered before. 

The outline of this paper is organized as follows. We introduce in Section \ref{sec:RI_oracle_overlap} some notations and show that the optimal (oracle) RI estimator involves the overlaps between the eigenvectors of 
the signal matrix $\C$ and its noisy estimate $\M$. In Section \ref{sec:RI_resolvent}, we observe that a convergence result on the resolvent of $\M$ not only gives us all the information about the eigenvalues, but also the eigenvectors. 
After motivating the study of the resolvent of the measurement matrix $\M$, we provide in Section \ref{section3} explicit expressions for three different perturbation processes. The first one is the case where we add a noisy matrix that is free with respect to the signal $\C$. The second model concerns multiplicative perturbations and includes 
the sample covariance matrix of (elliptically distributed) random variables. We also reconsider the case of the so-called `Information-Plus-Noise' matrix that deals with sample covariance matrices constructed from rectangular 
Gaussian matrices with an external source. 
The evaluation of the resolvent for each model is based on the powerful but non-rigorous replica method (which has been extremely successful in various contexts, including RMT or disordered systems-- see \cite{mezard1987spin}, or \cite{morone2014replica} for a more recent review). We will see that the derivation of our results using replicas can be done without too much effort and one can certainly imagine that our results can be proven rigorously, as was done in \cite{knowles2014anisotropic} for the resolvent or 
\cite{ledoit2011eigenvectors,allez2013eigenvectors,allez2014eigenvectors} for the overlaps of covariance matrices
and Gaussian matrices with external sources. We relegate all these technicalities in various appendices and only give our final results and their numerical verifications in Section \ref{section3}. Note in passing that we obtain using replicas the multiplication law of the $\str$-transform for product of free matrices (see Appendix \ref{sec:appendix_mult}), a derivation that we have not seen in the literature before. In Section \ref{section4}, we come back to the problem of statistical inference and apply the results of Section \ref{section3} to derive the optimal RIE for each considered model. In the multiplicative case, we recover and generalize the estimator recently derived by Ledoit and 
P{\'e}ch{\'e} for covariance matrices \cite{ledoit2011eigenvectors}. Each estimator is illustrated by numerical simulations, and we also provide some analytical formulas that can be of particular interest for real life problems. We then conclude this work with some open problems and possible applications of our results. 

\textbf{Conventions.} We use bold capital letters for matrices and bold lowercase letters for vectors. We denote usual RMT spectral transforms with calligraphic font. Finally, all acronyms and notations are summarized in Appendix \ref{app:abbreviations}.

\section{Rotationally Invariant Estimators, Eigenvector Overlaps and the Resolvent}\label{section2}

\subsection{The oracle estimator and the overlaps}
\label{sec:RI_oracle_overlap}

Throughout this work, we will consider the signal matrix $\C$ to be a symmetric matrix of dimension $N$ with $N$ that goes to infinity. We denote by $c_1 \ge c_2 \ge \dots \ge c_N$ its eigenvalues and by $|\b v_1 \rangle, | \b v_2 \rangle, \dots , | \b v_N \rangle$ their corresponding eigenvectors. The perturbed matrix $\M$ will be assumed to be symmetric with eigenvalues denoted by $\lambda_1 \ge \lambda_2 \ge \dots \ge \lambda_N$ associated to the eigenvectors $|\b u_1 \rangle, | \b u_2 \rangle, \dots , | \b u_N \rangle$. In the limit of large dimension, it is often more convenient to index the eigenvectors of both matrices by their corresponding eigenvalues, \ie $| \b u_i \rangle \to | \b u_{\lambda_i} \rangle$ and $| \b v_{i} \rangle \to | \b v_{c_i} \rangle$ for any integer $1 \le i \le N$, and this is the convention that we adopt henceforth.

We now attempt to construct an optimal estimator $\widehat{\b\Xi}(\M)$ of the true signal $\C$ that relies on the given dataset $\M$ at our disposal. We recall that our main assumption is that we have no prior insights on the eigenvectors of the matrix $\C$ so that all estimators considered below satisfy Eq.\ \eqref{RIH}. We conclude from the seminal work of \cite{takemura1983orthogonally} that any Rotational Invariant Estimator $\b\Xi(\M)$ of $\C$ shares the same eigenbasis as $\M$, that is to say
\begin{equation}
\label{best-RI-estim}
\b\Xi(\M) = \sum_{i=1}^{N} \xi_i \ket{\b u_i}\bra{\b u_i},
\end{equation}
where the eigenvalues $\xi_1, \dots, \xi_N$ are the quantities we wish to estimate. Next, the optimality of an estimator is defined with respect to a specific loss function (\eg the distance) and a standard metric is to consider the (squared) Euclidean (or Frobenius) norm that we shall denote by
\begin{align*}
\normb{ \C - \b\Xi(\M) }_{\mathbb{L}_2} \;\deq\; \Tr \left[ (\C - {\b\Xi}(\M) )^2\right] \,,
\end{align*} 
for a given RIE $\b\Xi(\M)$.
The best estimator with respect to this loss function is the solution of the following minimization problem 
\begin{align}\label{optimization-pb}
\widehat{\Xi}(\M) = \underset{\text{RI} \,\Xi(\M)}{\argmin} \,\normb{ \C - \b\Xi(\M) }_{\mathbb{L}_2}
\end{align}
considered over the set of all possible RI estimators $\b\Xi(\M)$. 
Since the only free variables in the constrained optimization problem \eqref{optimization-pb}
are the eigenvalues of $\b\Xi(\M)$,
it is easy to find the optimal solution: 
\begin{equation}
\label{oracle_estimator}
\widehat{\b\Xi}(\M) = \sum_{j=1}^{N} \wh\xi_i \ket{\b u_i} \bra{\b u_i}, \quad \widehat{\xi}_i = \sum_{j = 1}^{N} \braket{\b u_i}{\b v_j}^2 c_j\,,
\end{equation}
where we see that the optimal eigenvalues $\widehat{\xi}_i$ depend on the overlaps between 
the perturbed $\ket{\b u_i}$, the non-perturbed eigenvectors $\ket{\b v_j}$ and the eigenvalues of the
true matrix $\C$. 
A few comments on this estimator are in order. First, the estimator $\widehat{\xi}_i$ is designed 
to construct the best RI estimator $\widehat{\b\Xi}(\M)$ given in \eqref{best-RI-estim}. The consequence is that if we restrict our estimator to have the eigenvectors of the noisy matrix $\M$, then the naive approach that consists in \textit{substituting}\footnote{Remember that we have ranked the eigenvalues.} the eigenvalues $\{\widehat{\xi}_i\}_{i=1}^{N}$ with the true ones $\{c_i\}_{i=1}^{N}$ yields to a spectrum that is too wide. Indeed, it is not hard to see from \eqref{oracle_estimator} that the top eigenvalues are shrunk downward while the bottom ones are shrunk upward. In other words, the empirical spectral density (ESD) of the $\widehat{\xi}_i$ is narrower than the true one which shows that the RI estimator cannot be attained by the ``eigenvalues substitution'' procedure  independently proposed in \cite{bouchaud2009financial, el2008spectrum}, aside from the trivial case $\C = \In$.

We shall also see that the estimator $\widehat{\xi}_i$ is self-averaging in the 
large $N$-limit (in the sense that it converges almost surely, see Section \ref{sec:RI_resolvent}) 
and can thus be approximated by its expectation value 
\begin{align*}
\widehat{\xi}_i\approx \sum_{j = 1}^{N} \mathbb{E}\left[\braket{\b u_i}{\b v_j}^2\right] c_j.
\end{align*}
where $\mathbb{E}\left[\cdot \right]$ denotes the expectation value with respect to the random eigenvectors $(\ket{\b u_i})_i$
of the matrix $\M$. 
We will often use the following notation for the (rescaled) mean square overlaps
\begin{equation}
\label{overlaps}
O(\lambda_i, c_j) :=N \mathbb{E}\bigl[\braket{\b u_i}{\b v_j}^2\bigl]\,.
\end{equation}
Eqs.\ \eqref{oracle_estimator} and \eqref{overlaps} 
are the quantities of interest in this paper. In Statistics, the optimal solution \eqref{oracle_estimator} is sometimes called the \textit{oracle} estimator because it depends explicitly on the knowledge of the true signal $\C$. The ``miracle'' is that in the large $N$ limit, and for a 
large class of problems, one can actually express this oracle estimator in terms of the (observable) limiting spectral density (LSD) of $\M$ only.

\subsection{Relation between the resolvent and the overlaps}
\label{sec:RI_resolvent}

A convenient way to work out the overlap (\ref{overlaps}) is to study the resolvent of $\M$, defined as 
\begin{equation}
	{\G}_{\M}(z) \;\deq\; (z \In - \M)^{-1}.
\end{equation} 
The claim is that for $z$ not too close to the 
real axis, the matrix ${\G}_{\M}(z)$ is {\it self-averaging} in the large $N$ limit so that its value is independent of the specific realization of $\M$. More precisely, it means that ${\G}_{\M}(z)$ 
converges to a deterministic matrix 
for any fixed value (i.e. independent of $N$) of $z \in \mathbb{C}\setminus\mathbb{R}$ when $N\to \infty$. We will refer to this deterministic limit as the global law of ${\G}_{\M}(z)$ in the following.

The relation between the resolvent and the overlaps $O(\lambda_j, c_i)$ is relatively straightforward. For 
$z = \lambda - \ii \eta$ with $\lambda\in\mathbb{R}$ and $\eta \gg N^{-1}$, we have
\begin{align*}
{\G}_{\M}(\lambda - \ii \eta) = \sum_{k=1}^{N} \left[ \frac{\lambda + \ii \eta}{(\lambda - \lambda_k)^2 + \eta^2} \right] \ket{\b u_k}\bra{\b u_k}\,.
\end{align*}
If we take the trace of the above quantity, and take the limit $\eta \to 0$ (after $N\to \infty$), 
one obtains the limiting ``density of states'' (\ie the LSD) $\rho_{\M}$ (see Appendix A):
\begin{equation}
\Im \stj_{\M} (\lambda - \ii \eta) \equiv \Im \frac1N \Tr \qb{{\G}_{\M}(\lambda - \ii \eta)} = \pi \, \rho_{\M}(\lambda)\,.
\end{equation}
Similarly , the elements of $\Im {\G}_{\M}(\lambda - \ii \eta)$ can be written
for $\eta>0$ as
\begin{align}
\innerb{\b v_i}{\Im {\G}_{\M}(\lambda - \ii \eta)}{ \b v_i} \;=\; \sum_{k=1}^N \frac{\eta}{(\lambda-\lambda_k)^2+ \eta^2} 
\braket{\b v_i}{\b u_k}^2 \,.
\end{align}
This latter quantity is also self-averaging in the large $N$ limit (the overlaps $\braket{\b v_i}{\b u_k}^2, k=1,\dots,N$ with 
$i$ fixed display asymptotic independence when $N\to \infty$ so that the law of large number applies here) 
and we have 
\begin{align*}
&\innerb{\b v_i}{\Im {\G}_{\M}(\lambda - \ii \eta)}{ \b v_i} \; 
\\ & \underset{N\to \infty}{\rightarrow}\; 
 \;\;\int_\mathbb{R} \frac{\eta}{(\lambda-\mu)^2+\eta^2} O(\mu,c_i) \rho_{\M}(\mu) d\mu\,.  
\end{align*}
where the overlap function $O(\mu,c_i)$ is extended (continuously) to arbitrary values of $\mu$ inside the support of 
$\rho_{\M}$ in the large $N$ limit. 
Sending $\eta\to 0$ in this latter equation, we finally obtain the following formula valid in the large $N$ limit
\begin{align} 
\innerb{\b v_i}{\Im {\G}_{\M}(\lambda - \ii \eta)}{ \b v_i} \;\approx\;  \pi \rho_{\M}(\lambda) O(\lambda,c_i). \label{mean_squared_overlap}
\end{align}
Eq. (\ref{mean_squared_overlap}) will thus enable us to investigate the overlaps $O(\lambda, c_i)$ in great details through the calculation of the  elements of the resolvent 
${\G}_{\M}(z)$. This is what we aim for in the next section. We emphasize that the different equations of the mean square overlaps $O(\lambda,c_i)$ below will be expressed in the basis 
where $\C$ is diagonal without loss of generality (see Appendix \ref{appendixB} for more details). 

\section{Overlaps: Some Exact Results}\label{section3}

\subsection{Free additive noise}
\label{sec:overlaps_add}

The first model of noisy measurement that we consider is the case where the true signal $\C$ is corrupted by a free additive noise, that is to say 
\begin{equation}
\label{free_additive_noise}
\M = \C + \O {\bf B} \O^{\dag},
\end{equation}
where ${\bf B}$ is a fixed matrix with eigenvalues $b_1 > b_2 > \dots > b_N$ with limiting spectral 
density $\rho_{\B}$ and $\O$ is a random matrix chosen uniformly in the Orthogonal group $O(N)$ 
(i.e. according to the Haar measure). This family of models has found several applications in statistical physics of disordered systems subject to an external perturbation where the matrix $\M$ is interpreted as the Hamiltonian of the system, given by the sum of a deterministic term and a random term \cite{brezin1994correlation}. 
A simple example is when the noisy matrix $\O {\bf B} \O^{\dag}$ is a symmetric Gaussian random matrix with
independent and identically distributed (i.i.d.) entries, corresponding
to the so-called GOE (Gaussian Orthogonal Ensemble). By construction, the eigenvectors of a GOE matrix are invariant under rotation. 

It is now well known that the spectral density of $\M$ can be obtained from that of $\C$ and $\B$ using free addition, see \cite{voiculescu1992free} and, in the
language of statistical physics, \cite{zee1996law}. The statistics of the eigenvalues of $\M$ has therefore been investigated in great details, see \cite{brezin1995universala} and \cite{brezin1995universalb} for instance. However, the question of the eigenvectors has been much less studied, except 
recently in \cite{allez2013eigenvectors,allez2014eigenvectors} in the special case where $\O {\bf B} \O^{\dag}$ belongs to the GOE (see below).

For a general free additive noise, we show in Appendix B-2 that the global law estimate for the resolvent reads
in the large $N$ limit: 
\begin{equation}
\label{global_law_addition}
\avg{{\G}_{\M}(z)} = {\G}_{\C}(Z(z))
\end{equation}
where the function $Z(z)$ is given by 
\begin{equation}
\label{Z_addition}
Z(z) = z - \rtr_{\B}(\stj_{\M}(z)),
\end{equation}
and $\rtr_{\B}$ is the so-called $\rtr$-transform of $\B$ (see Appendix \ref{sec:appendix_transform} for a reminder of the definition of the different useful spectral transforms). 

Note that Eq.\ \eqref{global_law_addition} is a matrix relation, that simplifies when written in the basis where $\C$ is diagonal, since
in this case ${\G}_{\C}(Z)$ is also diagonal. Therefore, the evaluation of the overlap $O(\lambda, c)$ is straightforward using Eq.\ (\ref{mean_squared_overlap}). Let us define the Hilbert transform $\hil_{\M}(\lambda)$ which is simply the real part of the Stieltjes transform $\stj_{\M}(\lambda -  \ii\eta)$ in the limit $\eta \rightarrow 0$. Then the overlap for the free additive noise is given by:
\begin{equation}
\label{overlap_addition}
O(\lambda, c) = \frac{\beta_1(\lambda)}{(\lambda - c - \alpha_1(\lambda))^2 + \pi^2 \beta_1(\lambda)^2 \rho_{\M}(\lambda)^2},
\end{equation}
where $c$ is the corresponding eigenvalue of the unperturbed matrix $\C$, and where we defined: 
\begin{equation}
\label{decomposition_R}
\begin{dcases}
\alpha_1(\lambda) :=  \Re[\rtr_{\B} \left( \hil_{\M}(\lambda) + \ii \pi \rho_{\M}(\lambda) \right)],  \\
\beta_1(\lambda) := \frac{\Im[\rtr_{\B} \left( \hil_{\M}(\lambda) + \ii \pi \rho_{\M}(\lambda) \right)]}{\pi \rho_{\M}(\lambda)}\,.
\end{dcases}
\end{equation}

As a first check of these results, let us consider the normalized trace of Eq. (\ref{global_law_addition}) and then set $u=\stj_{\M}(z) = \stj_{\C}(Z(z))$. One can find by using the Blue transform, defined in \eqref{app_blue}, that we indeed retrieve the free addition formula $\rtr_{\M}(u) = \rtr_{\C}(u) + \rtr_{{\bf B}}(u)$ when $N\to \infty$, as it should be. 

\subsubsection{Deformed GOE}

\begin{figure}
   \includegraphics[scale = 0.32]{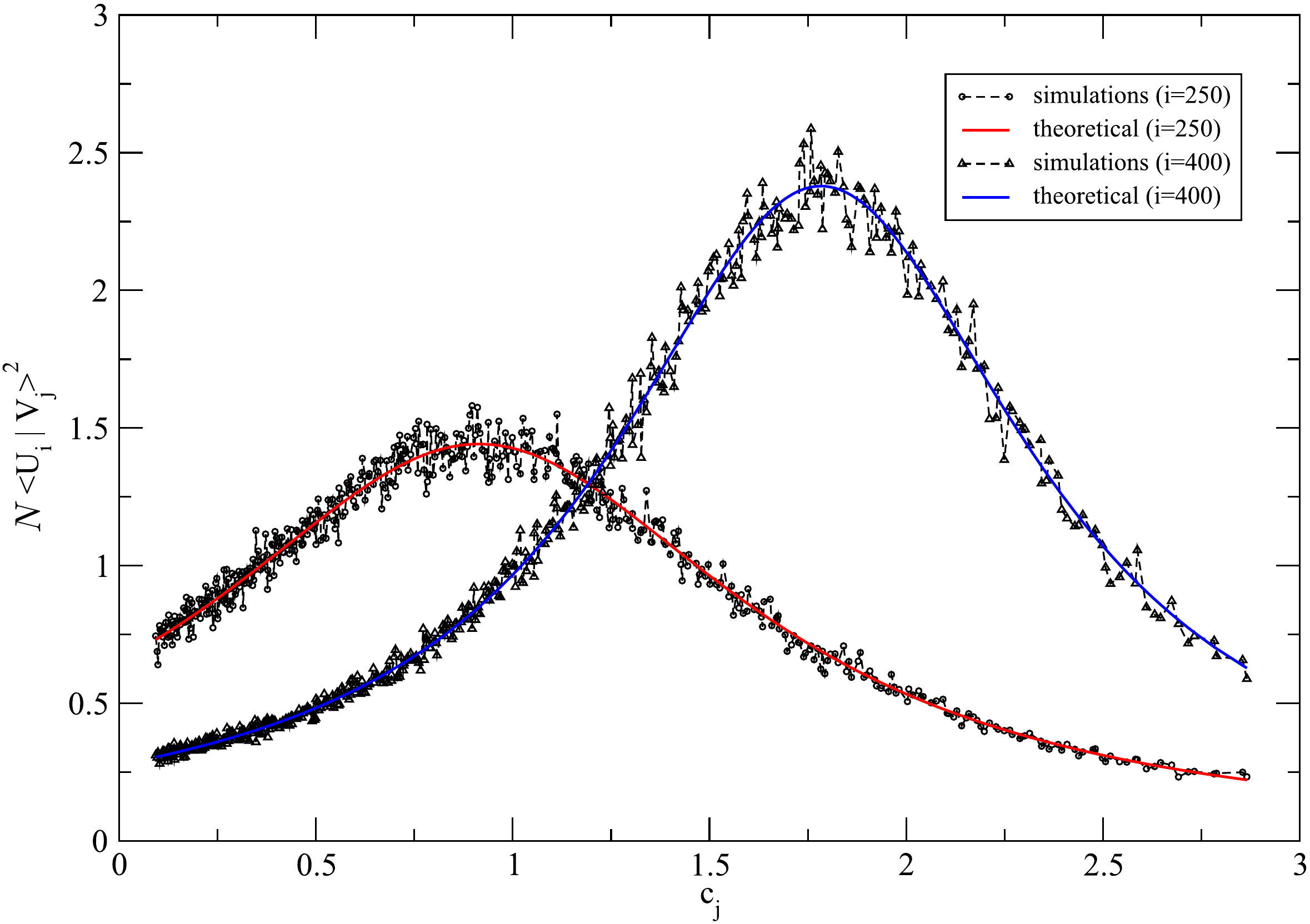} 
   \caption{Computations of the rescaled overlap $O(\lambda, c)$ as a function of $c$ in the free addition perturbation. 
   We chose $i = 250$, $\C$ a Wishart matrix with parameter $q = 0.5$ and $\B$ a Wigner matrix with $\sigma^2 = 1$. The black dotted points are computed using numerical simulations and the plain red curve is the theoretical predictions Eq. \eqref{overlap_addition}. The agreement is excellent. For $i = 250$, we have $c_i \approx 0.83$ and we see that the peak of the curve is in that region. The same observation holds for $i = 400$ where $c_i \approx 1.66$. {
   The numerical curves display the empirical mean values of the overlaps 
   over 1000 samples of $\M$ given by Eq. \eqref{free_additive_noise} with $\C$ fixed}.}
   \label{Overlaps_add}
\end{figure}

As a second verification, we specialize our result to the case where $\O\BBB \O^{\dag}$ is a GOE matrix such that the entries have a variance equal to $\sigma^2/N$. Then, one has $\rtr_{\B}(z) = \sigma^2 z$ meaning that Eq. \eqref{Z_addition} simply becomes $Z(z) = z - \sigma^2 \stj_{\M}(z)$. This allows us to get a simpler expression for the overlap:
\begin{equation}
\label{overlap_addition_gaussian}
O(\lambda, c)= \frac{\sigma^2}{(c -\lambda + \sigma^2 \hil_{\M}(\lambda))^2 + \sigma^4 \pi^2 \rho_{\M}(\lambda)^2},
\end{equation}
which is exactly the result derived in \cite{allez2013eigenvectors,allez2014eigenvectors} using other methods. In Fig.\ \ref{Overlaps_add}, we illustrate this formula in the case where $\C$ is an isotropic Wishart matrix of parameter $q$, by taking \eg $\C = T^{-1} \HHH \HHH^{\dag}$ where $\HHH$ is a symmetric matrix of size 
$N \times T$ filled with i.i.d. standard Gaussian entries and $q = N/T$. We set $N = 500$, $T = 1000$, and take $\O{\bf B}\O^{\dag}$ as a GOE matrix with variance $1/N$. For a fixed $\C$, we generate 1000 samples of $\M$ given by Eq. (\ref{free_additive_noise}) for which we can measure numerically the overlap quantity. We see that the theoretical prediction (\ref{overlap_addition_gaussian}) agrees remarkably with the numerical simulations.

\subsection{Free multiplicative noise and empirical covariance matrices}
\label{sec:overlaps_mult}

Our second model deals with {\it multiplicative} noise in the following sense: we consider that the noisy measurement matrix $\M$ can be written as
\begin{equation}
\label{free_multiplicative_noise}
\M = \sqrt{\C}  \O {\bf B} \O^{\dag} \sqrt{\C},
\end{equation}
where again $\C$ is the signal, ${\bf B}$ is a fixed matrix with eigenvalues $b_1 > b_2 > \dots > b_N$ with 
limiting density $\rho_{\B}$ and $\O$ is a random matrix chosen in the Orthogonal group $O(N)$ according to the Haar measure. Note that we implicitly requires that $\C$ is positive definite with Eq. (\ref{free_multiplicative_noise}), 
so that the square root of $\C$ is well defined. 

An explicit example of such a problem is provided by sample covariance matrices where ${\bf B}$ is a Wishart matrix \cite{wishart1928generalised}), which is of particular interest in multivariate statistical analysis.  We shall come back later to this application. The Replica analysis leads to the following systems of equations (see Appendix \ref{sec:appendix_mult}) for the general problem of a free multiplicative noise above, Eq. \eqref{free_multiplicative_noise}:
\begin{equation}
\label{global_law_multiplication}
z \avg{{\G}_{\M}(z)} = Z(z) {\G}_{\C}(Z(z)),
\end{equation}
with:
\begin{equation}
\label{Z_multiplication}
Z(z) = z \str_{\B}(z\stj_{\M}(z) - 1),
\end{equation}
where $\str_{\B}$ is the so-called $\str$-transform of $\B$ (see Appendix \ref{sec:appendix_transform}) 
and $\stj_{\M}$ is the normalized trace of ${\G}_{\M}(z)$. The latter obeys, from Eq.\ (\ref{global_law_multiplication}), the self-consistent equation:
\begin{equation}
\label{generalized_MP}
z \stj_{\M}(z) = Z(z) \stj_{\C}(Z(z)).
\end{equation}
Again, Eq. \eqref{global_law_multiplication} is a matrix relation, that simplifies when written in the basis where $\C$ is diagonal.
Note that Eqs \eqref{generalized_MP} and \eqref{Z_multiplication} allow us to retrieve the usual free multiplicative convolution, that is to say:
\begin{equation}
\str_{\M}(u) = \str_{\C}(u) \str_{\B}(u).
\end{equation}
This result is thus the analog of our result \eqref{global_law_addition} in the multiplicative case. We refer the reader to the appendix \ref{sec:appendix_mult} for more details. We emphasize that for technical reasons, we restrict $\B$ to have a normalized trace that differs from zero.

With the global law estimate for the resolvent given by Eqs. \eqref{global_law_multiplication} and \eqref{Z_multiplication} above, we can obtain a 
general overlap formula for the free multiplicative noise case. Let us define the following functions
\begin{equation}
\label{decomposition_S}
\begin{dcases}
\alpha_{2}(\lambda) :=  \underset{z \rightarrow \lambda - i0^{+}}{\lim} \Re\left[ \frac{1}{\str_{\B}(z \stj_{\M}(z) - 1)} \right]  \\
\beta_{2}(\lambda) :=  \underset{z \rightarrow \lambda - i0^{+}}{\lim} \Im\left[ \frac{1}{\str_{\B}(z \stj_{\M}(z) - 1)} \right] \frac{1}{\pi \rho_{\M}(\lambda)},
\end{dcases}
\end{equation}
then the overlap $O(\lambda, c)$ between the eigenvectors of $\C$ and $\M$ are given by:
\begin{equation}
\label{anisotropic_correlation_overlap}
O(\lambda, c) = \frac{c \beta_{2}(\lambda)}{(\lambda - c \alpha_{2}(\lambda))^2 + \pi^2 c^2 \beta_{2}(\lambda)^2 \rho_{\M}(\lambda)^2}.
\end{equation}
In order to give more insights on our results, we will now specify these results to some well-known applications of multiplicative models in RMT. 

\subsubsection{Empirical covariance matrix}

As mentioned previously, the most famous application of a model of the form (\ref{free_multiplicative_noise}) is given by the sample covariance estimator that we recall briefly. Let us define by $\R \deq (R_{it}) \in \mathbb{R}^{N \times T}$ the observation matrix whose columns represent the collected samples of size $T$ that we assume to be independently and identically distributed with zero mean. The $N$ elements of each sample generally display 
some degree of interdependence, that is often represented by the \textit{true} (or also \textit{population}) covariance matrix $\C \deq (C_{ij}) \in \mathbb{R}^{N\times N}$, defined as $\langle R_{it} R_{jt'} \rangle = C_{ij} \delta_{t,t'}$, where $\delta_{t,t'}$ is the Kronecker symbol.  As the signal $\C$ is unknown, the classical way to estimate the covariances is to compute the empirical (or sample) covariance matrix thanks to Pearson estimator
\begin{equation}
\label{eq:empirical_covariance}
\M \;\deq\; \frac{1}{T} \R \R^{\dag} \equiv \sqrt{\C} \X \X^{\dag} \sqrt{\C}\,,
\end{equation}
where $\X$ is a $N \times T$ random matrix where all elements are i.i.d. random variables with zero mean and variance $T^{-1}$. It is easy to see that this model is a particular case of the model \eqref{free_multiplicative_noise} where 
$\B \equiv \X \X^{\dag}$ whose $\str$-transform has an explicit expression \cite{tulino2004random}:
\begin{equation}
\label{eq:s_tranform_wishart}
\str_{\B}(x)= \frac{1}{1+qx}, \qquad q= \frac{N}{T}\,.
\end{equation}
Using our general results Eqs. \eqref{global_law_multiplication} and \eqref{Z_multiplication}, we obtain
\begin{equation}
\label{MP_burda}
\begin{cases}
z \avg{{\G}_{\M}(z)} = Z(z) {\G}_{\C}(Z(z))\,, \\
\quad Z(z) \;\deq\; \frac{z}{1-q+qz\stj_{\M}(z)}\,.
\end{cases}
\end{equation}
which is exactly the result found in \cite{burda2004signal} and also in \cite{knowles2014anisotropic} at leading order. We can therefore recover the well-known Mar{\v c}enko-Pastur equation \cite{marchenko1967distribution} which gives a fixed point equation satisfied by the Stieltjes transform of $\M$ in term of the Stieltjes transform of the true matrix $\C$
\begin{equation}
\label{MP_equation}
\begin{cases}
z\stj_{\M}(z) = Z(z)\stj_{\C}(Z(z))\,, \\
\quad Z(z) \;\deq\; \frac{z}{1-q+qz\stj_{\M}(z)}\,.
\end{cases}
\end{equation}
The expression of the limiting overlaps can be further simplified in this particular case to
\begin{align}
\label{MP_overlap}
& O(\lambda, c) = \nonumber \\
& \frac{q c \lambda}{(c (1-q) - \lambda + q c \lambda \hil_{\M}(\lambda))^2 + q^2 \lambda^2 c^2 \pi^2 \rho_{\M}(\lambda)^2}\,
\end{align}
and we recover, as expected, the Ledoit $\&$ P\'ech\'e result established in \cite{ledoit2011eigenvectors}. As a conclusion, 
our result generalizes the standard Mar{\v c}enko $\&$ Pastur formalism to an arbitrary multiplicative noise term $\O\B \O^{\dag}$.

\subsubsection{Elliptical ensemble}

A slightly more general application of the model (\ref{free_multiplicative_noise}) is when we suppose that the entries of the observation matrix $\R$ can be written as the product of two independent sources $R_{it} = \sigma_{t} Y_{it}$. The $\{Y_{it}\}$ are characterized by the true signal, i.e.\ $C_{ij} = \langle Y_{it} Y_{jt'} \rangle \delta_{t,t'}$ and are generated independently from the same distribution at time $t$ that will be assumed to be Gaussian in our case. The $\{\sigma_t\}$ are such that $\langle \sigma^2 \rangle = 1$ and allows to add a time-dependent volatility with a factor $\sigma_t$ that is common to all variables at time $t$. This defines the class of elliptical distributions and the most famous application is when the $\{\sigma_t\}$ are drawn from a inverse-gamma distribution which leads to the multivariate Student distribution \cite{bouchaud2003theory} (see Sec. \eqref{sec:oracle_mult_elliptical} below). The corresponding empirical correlation matrix can be written as
\begin{equation}
\label{covariance_elliptic}
\M =   \sqrt{\C} \X \Sigma \X^{\dag} \sqrt{\C},
\end{equation}
where $\Sigma := \text{diag}(\sigma_1^2, \sigma_2^2, \dots, \sigma_T^2)$ and $\X$ is defined as in Eq.\ \eqref{eq:empirical_covariance}. This model has been subject to several studies in RMT, see e.g. \cite{biroli2007student} \cite{burda2005spectral}, \cite{zhang2006spectral} or \cite{el2009concentration}. In all these works, the expression of the limiting Stieltjes transform of the 
spectral density is quite complex, except for the case where $\C$ is the identity matrix. We find here that we can in fact obtain a self-consistent expression for the global law estimate of the corresponding resolvent by introducing the appropriate transforms. Our result generalizes the time-independent result of \cite{burda2004signal} or \cite{knowles2014anisotropic}, and also provides a tractable equation for the limiting eigenvalues density. 

Before stating the result for the elliptical model \eqref{covariance_elliptic}, one has to be careful with the $\str$-transform of $\B$. Indeed, it is in fact more convenient to work with the ``dual'' matrix $\B_* := \sqrt{\Sigma} \X^{\dag} \C \X \sqrt{\Sigma}$ 
in order to use the free multiplication formula. We then obtain the $\str$-transform of $\B_*$ to finally express  
the Stieltjes transform of $\stj_{\B_*}$ as a function $\stj_{\B}$, 
simply by noticing that $\B_*$ has the same eigenvalues as $\B$ 
and the additional zero eigenvalue with multiplicity $T-N$. 
The final result reads, after elementary manipulations of the $\ttr$-transform, 
\begin{equation}
\label{S_dual_formula}
\str_{\B_*}(x) = \frac{x+1}{x+q} \str_{\B}\left(\frac{x}{q}\right).
\end{equation}
In a nutshell, applying the result Eq.\ (\ref{Z_multiplication}) to the elliptical case leads to the result
\begin{equation}
\label{Z_cov_elliptic}
Z(z) = \frac{z}{1-q+qz\stj_{\M}(z)} \str_{\Sigma}(q(z\stj_{\M}(z) - 1))
\end{equation}
with Eqs.\ \eqref{global_law_multiplication} and \eqref{generalized_MP} unchanged. With Eq.\ \eqref{Z_cov_elliptic}, 
the general result \eqref{generalized_MP}
extends the standard Mar{\v c}enko-Pastur formula to a  time-dependent\footnote{in the sense that the volatility depends on 
the observation time $t$.} framework. Note that the corresponding self-consistent equation for the Stieltjes transform $\stj_{\M}(z)$ has been obtained in previous studies \cite{burda2005spectral, zhang2006spectral, el2009concentration} yet stated in a different form. 
One can easily specialize the result of the overlaps $O(\lambda, c)$ to any $\lambda$ and $c$ as a function of the spectral measure of $\Sigma$. However, we do not find an expression as tractable as Eq. \eqref{MP_overlap}. 

\begin{figure}
   \includegraphics[scale = 0.32]{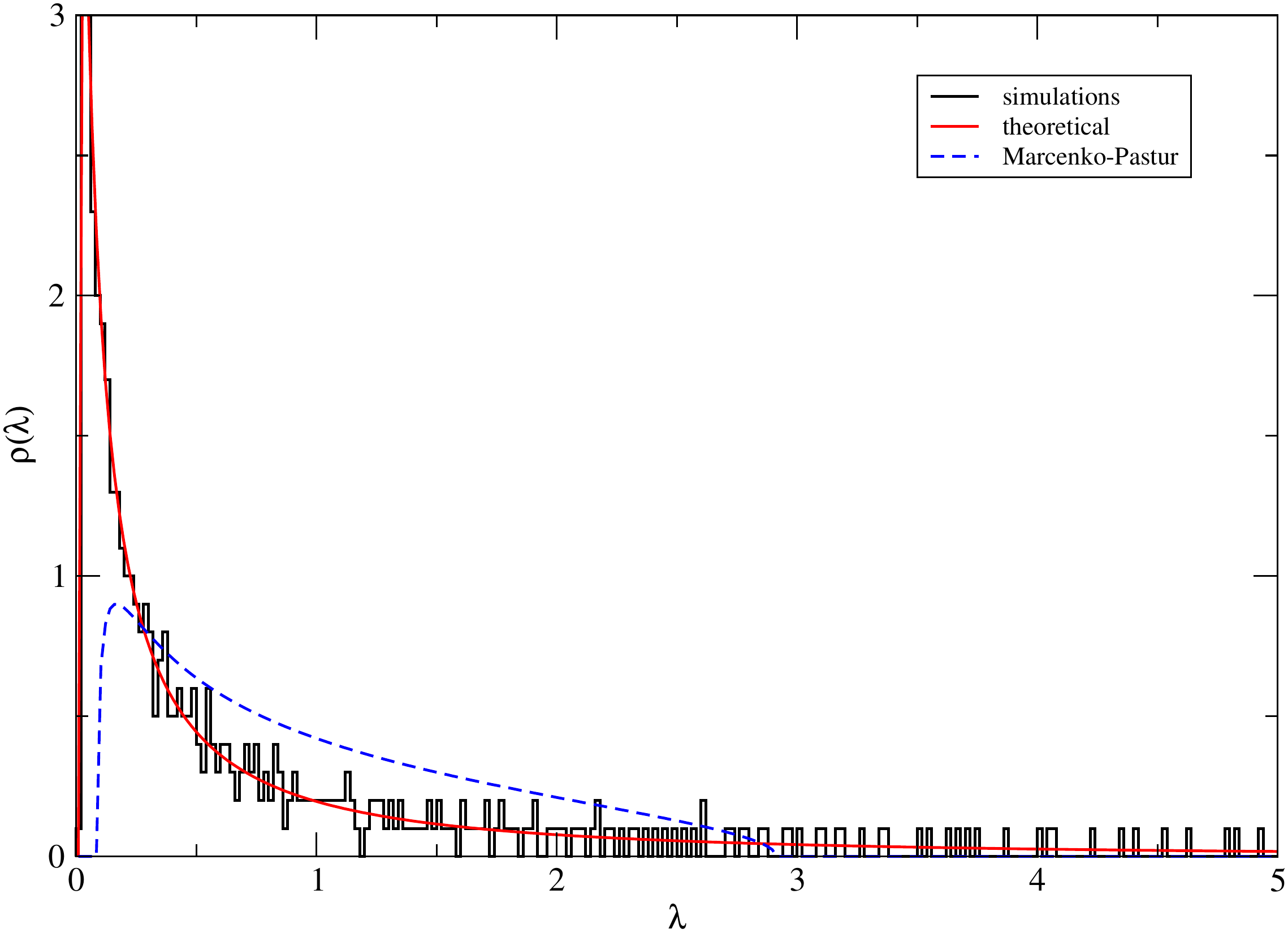} 
   \caption{Theoretical predictions of the density of states from Eq. (\ref{global_law_multiplication}) (red line) compared to simulated data when $\C$ is a $500 \times 500$ inverse-Wishart matrix (parameter $\kappa = 0.2$) and $\Sigma$ is a white Wishart with $q_0 = 0.6$. The agreement of our theoretical estimate is excellent and differs strongly from the classical Mar{\v c}enko-Pastur density (blue dotted curve) }
   \label{density_cov}
\end{figure}

\begin{figure}
   \includegraphics[scale = 0.32]{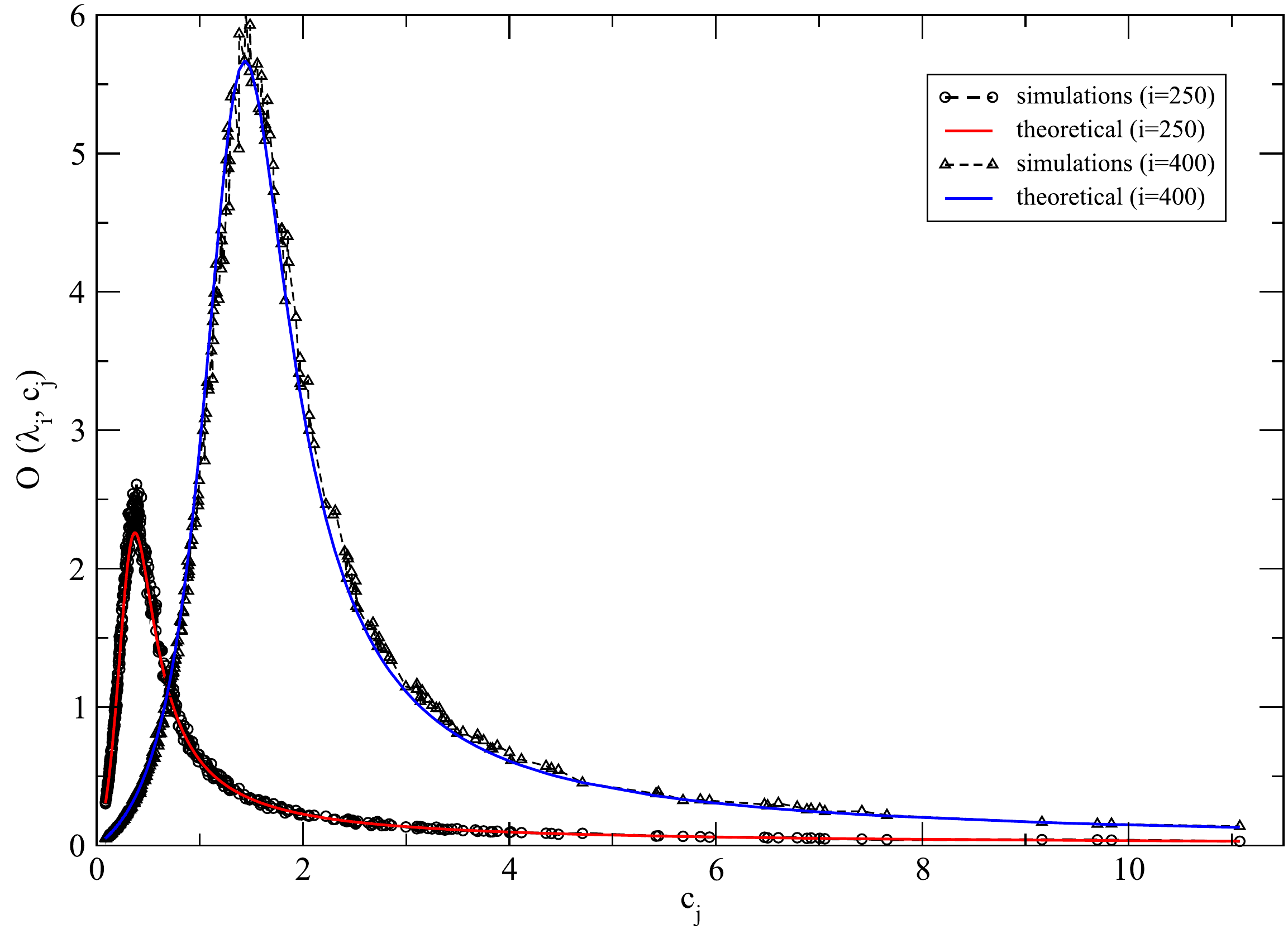} 
   \caption{Rescaled overlap $ O(\lambda, c)$ as a function of $c_j$ in the free multiplicative perturbation with $N = 500$. We chose $\C$ as an inverse-Wishart matrix with parameter $\kappa = 0.2$ and $\Sigma$ a Wishart matrix with $q_0 = 0.6$. The black dotted points are computed using numerical simulations and the plain curves are the theoretical predictions Eq. (\ref{anisotropic_correlation_overlap}). For $i = 250$ (resp. $i = 400$), we have $c_i \approx 0.37$ (resp. $c_i \approx 1.48$) and we see that the peak of the curve is in that region for both value of $i$. }
   \label{overlap_cov}
\end{figure}

Finally, let us now show that Eq. \eqref{generalized_MP} can be useful for practical purposes in order to construct non-trivial models. Suppose that $\C$ is an inverse-Wishart matrix (see Section IV.B. for the definition of this law) with parameter $\kappa = 0.2$ and define $\Sigma$ to be a 
Wishart matrix of size $T \times T$ and parameter $q_0 = 0.6$. We follow the same 
numerical procedure as in the free additive noise case. We compare in Fig. \ref{density_cov} our theoretical result Eq. \eqref{generalized_MP} with empirical simulations and the agreement is remarkable. The same conclusion holds for the overlap (see Fig. \ref{overlap_cov}).

Note that the results obtained in this section could be extended to the case where the diagonal matrix $\Sigma$ is not positive definite, for applications in regression analysis (see for example \cite{reigneron2011principal}). It can also be used for studying Maronna's robust estimators of $\C$ as the deterministic equivalent of such estimators in the large $N$ falls down into the model \eqref{free_multiplicative_noise} \cite{couillet2015random}.

\subsection{Information-Plus-Noise matrix}
\label{sec:overlaps_info}

The last model we will treat here is the so-called `Information-Plus-Noise' family of matrix \cite{dozier2007empirical}. In this model, 
we suppose that at each time $t$, we observe a $N$-dimensional vector whose entries are given by $R_{it} \deq A_{it} + \sigma X_{it}$ for any $i = 1,\dots,N$ where the signal is contained in the variable $A_{it}$ which is perturbed by an additive noise $\sigma X_{it}\in \mathbb{R}^N$. 
We will assume that the entries of $\X = (X_{it}) \in \mathbb{R}^{N\times T}$ are i.i.d. Gaussian random variables with zero mean and unit variance. In the case where the number of samples $T \gg N$, the empirical covariance matrix given by
\begin{equation}
\label{info_noise}
\M = \frac1T \R \R^{\dag} = \frac1T (\AAA+ \sigma \X)(\AAA + \sigma \X)^{\dag}
\end{equation}
is a good estimator of $\frac1T \AAA \AAA^{\dag} + \sigma^2 \In$. 
This model is of particular interest in signal processing, in order to detect the number of sources and their direction of arrival \cite{schmidt1986multiple}. Another example of application of this model comes from Finance where one may want to estimate the integrated covariance matrix from high-frequency noisy observation where the matrix $\X$ plays the role of the \textit{microstructure} noise \cite{xia2014integrated}. 

As usual, in the case where $T \sim {\cal O}(N)$, the empirical estimator cannot be fully trusted. The main assumption of the model is again the convergence of the empirical density of eigenvalues of $\C:=T^{-1} \AAA \AAA^{\dag}$ towards 
a limiting density $\rho_\C$. 
The global law of the Information-Plus-Noise matrix reads, in a matrix sense:
\begin{equation}
\label{global_law_info_noise}
\avg{{\G}_{\M}(z)} = \left( (z Z(z) - \sigma^2(1-q)) - Z(z)^{-1} \C \right)^{-1},
\end{equation}
where we have defined
\begin{equation}
\label{Z_info_noise}
Z(z) = 1 - q \sigma^2 \stj_{\M}(z).
\end{equation}
This global law result \eqref{global_law_info_noise} has already been obtained in a mathematical context by the authors of \cite{hachem2013bilinear} for applications in wireless communications and signal processing. However, it is satisfactory to see that the replica method is able to reproduce this result. If we take the normalized trace of the 
above equation, we find that the Stieltjes transform reads
\begin{align}
& \stj_{\M}(z) = \nonumber \\
& \int \frac{\dd c \rho_{\C}(c)}{z(1-q\sigma^2 \stj_{\M}(z)) - \sigma^2(1-q) - \frac{c}{1-q\sigma^2 \stj_{\M}(z)}},
\end{align}
which is the result obtained in \cite{dozier2007empirical}. As far as we understand, the authors of \cite{hachem2013bilinear} did not discuss the overlaps 
in the present context. The final expression for $O(\lambda, c)$ is quite cumbersome, but again completely explicit, 
and reads:
\begin{eqnarray}
\label{overlap_info_noise}
O(\lambda, c) & = & \frac{q \sigma^2 \alpha_3(\lambda) (\lambda \alpha_3(\lambda) + c)}{ \chi_1^2(\lambda,c) + \chi_2^2(\lambda,c)}  \\
\alpha_3(\lambda) & \deq & \zeta^2(\lambda) + q^2 \sigma^4 \pi^2 \rho_{\M}^2(\lambda) \nonumber \\
\chi_1(\lambda) & \deq & (\lambda \alpha_3(\lambda) - c) \zeta(\lambda) - \alpha_3(\lambda)\sigma^2(1-q) \nonumber \\
\chi_2(\lambda) & \deq & (\lambda \alpha_3(\lambda) + c) q \sigma \pi \rho_{\M}(\lambda) \nonumber \\
\zeta(\lambda) & \deq & 1-q\sigma^2 \hil_{\M}(\lambda)\,. \nonumber
\end{eqnarray}

\begin{figure}
   \includegraphics[scale = 0.32]{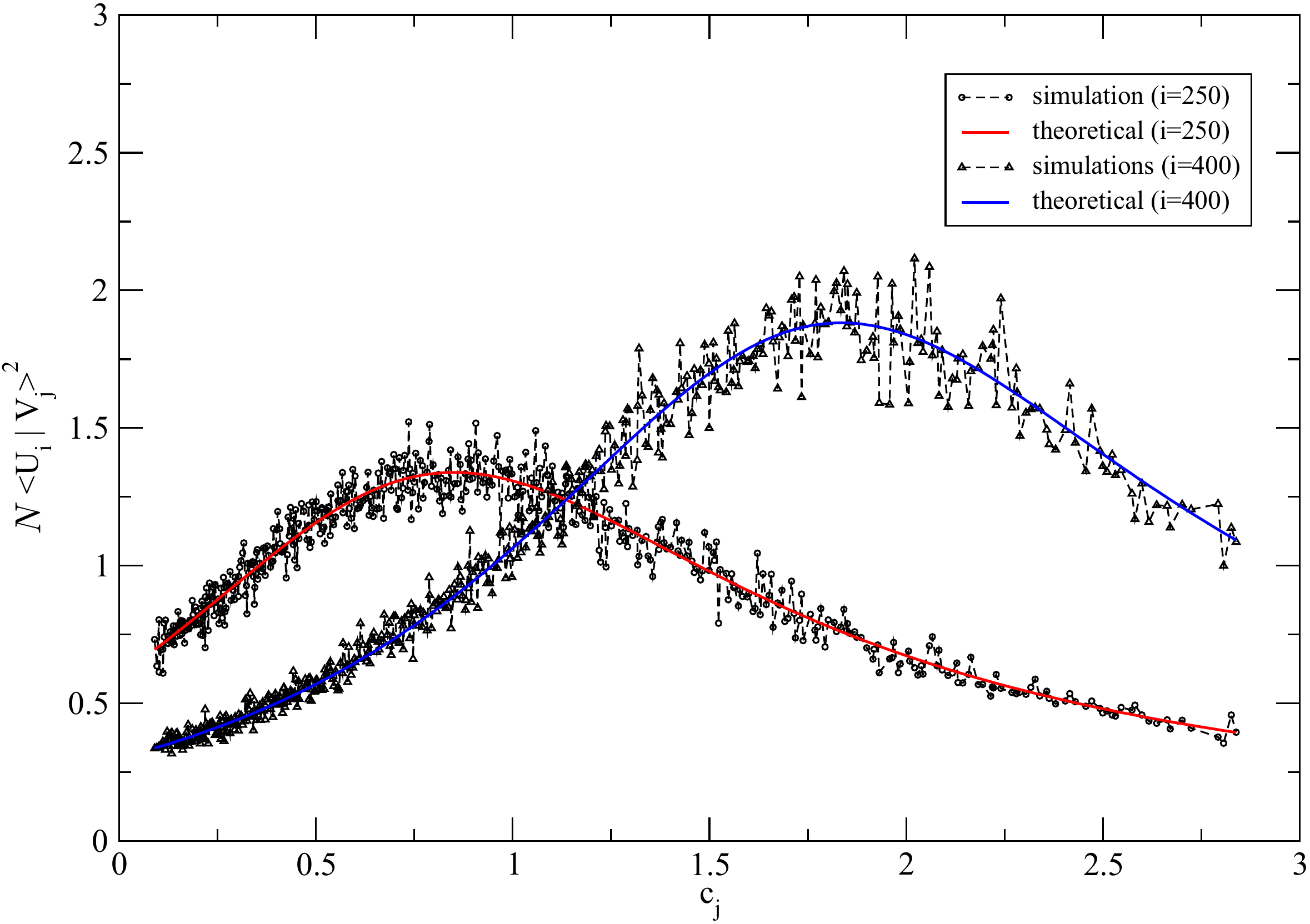} 
   \caption{Rescaled overlap $O(\lambda, c)$ as a function of $c_j$ in the information-plus-noise model with $N = 500$. We chose $\AAA$ and $\X$ to be a $N \times T$ Gaussian matrix and $T = 2N$. The black dotted points are computed using numerical simulations and the plain curves are the theoretical predictions Eq. (\ref{overlap_info_noise}). For $i = 250$ (resp. $i = 400$), we have $c_i \approx 0.83$ (resp. $c_i \approx 1.66$) and we see that the peak of the curve is in that region for both value of $i$. }
   \label{overlap_info}
\end{figure}

For the sake of completeness, we provide a numerical example for the overlap (\ref{overlap_info_noise}) where $\AAA$ is a Gaussian matrix of size $N \times T$ with $q = 0.5$ and $N = 500$ with variance 1. The perturbation is a Gaussian noise of same size with $\sigma = 1$. The procedure is the same than in the previous section and we give a numerical example in Fig. \ref{overlap_info}.

\section{Optimal rotational invariant estimator}\label{section4}

The above resolvent and overlap formulas for various models of random matrices are 
the central results of this study. Equipped with these results, we can now tackle the problem 
of the optimal RIE of the signal $\C$. As we shall see, the high-dimensional limit $N \to \infty$ allows 
one to reach some degree of universality.  First, we rewrite the RIE  (\eqref{oracle_estimator}) as:
\begin{equation*}
\widehat{\xi}_i \underset{N \rightarrow \infty}{=} \frac1N \sum_{j = 1}^{N} O(\lambda_i, c_j) c_j \approx \int c\, \rho_{\C}(c) \, O(\lambda_i,c) \, \dd c .
\end{equation*}
Quite remarkably, as we show below, the optimal RIE can be expressed, in the three above cases, as a function of the spectral measure of the observable (noisy) $\M$ only. 

Let us however stress that the nonlinear ``shrinkage'' estimators $\widehat{\xi}_i$ we obtain below are \textit{{a} priori} 
valid in the support of $\M$ only. An interesting problem for future research would be to extend the results obtained here for the bulk eigenvalues to  
the spiked eigenvalues, also called \textit{outliers}. Here, we assume that there are no spikes and we perform the optimal RIE for each models.  

\subsection{Free additive noise}
\label{sec:oracle_add}

We now specialize the RIE and we begin with the free additive noise case for which the noisy measurement is given by
\begin{equation*}
\M = \C + \O \B \O^{\dag}.
\end{equation*}
It is easy to see from Eqs. (\ref{mean_squared_overlap}) and (\ref{global_law_addition}) that:
\begin{eqnarray}
\widehat{\xi}_i & = & \frac{1}{\pi \rho_{\M}(\lambda_i)} 
\lim_{z \rightarrow \lambda_i - i0^{+}} \Im\qBB{\int \frac{c\, \rho_{\C}(c)}{Z(z) - c} \, \dd c} \nonumber \\
& = & \frac{1}{N\pi \rho_{\M}(\lambda_i)} \underset{z \rightarrow \lambda_i - i0^{+}}{\lim} \, \Im \Tr \left[ {\G}_{\M} (z) \C \right]\,, \nonumber \\
\end{eqnarray}
where $Z(z)$ is given by Eq. \eqref{Z_addition}. From Eq. \eqref{global_law_addition} one also has 
$\Tr [{\G}_{\M}(z) \C] = N (Z(z) \stj_{\M}(z) - 1)$, and using Eqs. \eqref{Z_addition} and \eqref{decomposition_R}, 
we end up with:
\begin{align*}
& \underset{z \rightarrow \lambda - i0^{+}}{\lim} \, \Im \Tr \left[ {\G}_{\M}(z) \C \right] \\
&\qquad\qquad = N \pi \rho_{M}(\lambda) \left[ \lambda - \alpha(\lambda) - \beta(\lambda) \hil_{\M}(\lambda) \right].
\end{align*}
We therefore find the following optimal RIE nonlinear ``shrinkage'' function $F_1$:
\begin{equation}
\label{oracle_free_add}
\widehat{\xi}_i = F_1(\lambda_i); \quad F_1(\lambda)= \lambda - \alpha_1(\lambda) - \beta_1(\lambda) \hil_{\M}(\lambda),
\end{equation}
where $\alpha_1, \beta_1$ are defined in Sect. III.A, Eq. (\ref{decomposition_R}). 
This result states that if we consider a model where the signal $\C$ is perturbed with an additive noise 
(that is free with respect to $\C$), 
the optimal way to 'clean' the eigenvalues of $\M$ in order to get $\widehat{\Xi}(\M)$ is to keep the eigenvectors of $\M$ and apply the nonlinear shrinkage formula \eqref{oracle_free_add}. We see that the non-observable oracle estimator converges in the limit $N \to \infty$ towards a deterministic function of the observable eigenvalues.

\subsubsection{Deformed GOE}

Let us consider the case where $\B$ is a GOE  matrix. Using the definition of $\alpha_1$ and $\beta_1$ given in Eq. (\ref{decomposition_R}), the nonlinear shrinkage function is given by
\begin{equation}
\label{oracle_gaussian}
F_1({\lambda}) = \lambda - 2 \sigma^2 \hil_{\M}(\lambda).
\end{equation}
Moreover, suppose that $\C$ is also a GOE matrix so that $\M$ is a also a GOE matrix with variance 
$\sigma^2_{\M} = \sigma^2_{\C} + \sigma^2$. As a consequence, the Hilbert transform of $\M$ can be computed straightforwardly from the Wigner semicircle law and we find
\begin{equation*}
\hil_{\M}(\lambda) = \frac{\lambda}{2\sigma_{\M}^2}\,.
\end{equation*}
The optimal cleaning scheme to apply in this case is then given by: 
\begin{equation}
\label{oracle_gaussian_Wigner}
F_1({\lambda})= \lambda \left( \frac{\sigma_{\C}^2}{\sigma^2_{\C} + \sigma^2} \right)\,,
\end{equation}
where one can see that the optimal cleaning is given by rescaling the empirical eigenvalues by the signal-to-noise ratio. This result is expected in the sense that we perturb a Gaussian signal by adding a Gaussian noise. We know in this case that the optimal estimator of the signal is given, element by element, by the Wiener filter \cite{wiener1949extrapolation}, and this is exactly the result that we have obtained with \eqref{oracle_gaussian_Wigner}. We can also notice that the ESD of the cleaned matrix is narrower than the true one. Indeed, let us define the signal-to-noise ratio $\text{SNR} = \sigma_{\C}^{2}/\sigma_{\M}^{2} \in [0,1]$, and it is obvious from \eqref{oracle_gaussian_Wigner} that $\widehat{\Xi}(\M)$ is a Wigner matrix with variance $\sigma_{\Xi}^{2} \times \text{SNR}$ which leads to 
\begin{equation}
\sigma_{\M}^{2} \ge \sigma_{\C}^{2} \ge \sigma_{\C}^{2} \times \text{SNR}\,,
\end{equation}
as it should be.

\begin{figure}
   \includegraphics[scale = 0.32]{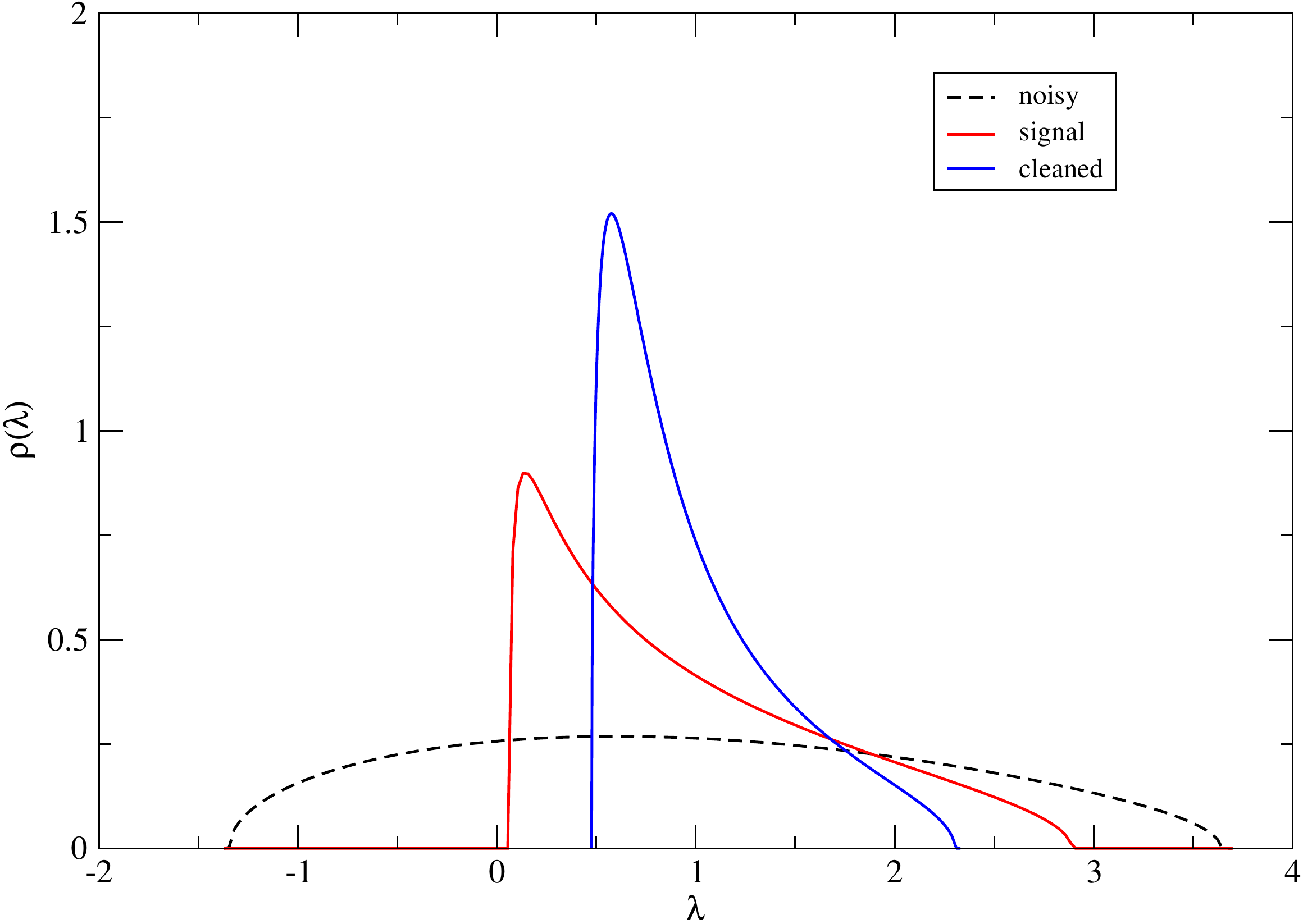} 
   \caption{Eigenvalues of the noisy measurement $\M$ (black dotted line) compared to the true signal $\C$ drawn from a $500 \times 500$ Wishart matrix of parameter $q_0 = 0.5$ (red line). We have corrupted the signal by adding a GOE matrix with radius 1. The eigenvalues density of $\M$ allows negative values while the true one has only positive values. The blue line is the LSD of the optimally cleaned matrix. We clearly notice that the cleaned eigenvalues are all positive and its spectrum is narrower than the true one, while preserving the trace.}
   \label{oracle_addition_density}
\end{figure}

\begin{figure}
   \includegraphics[scale = 0.32]{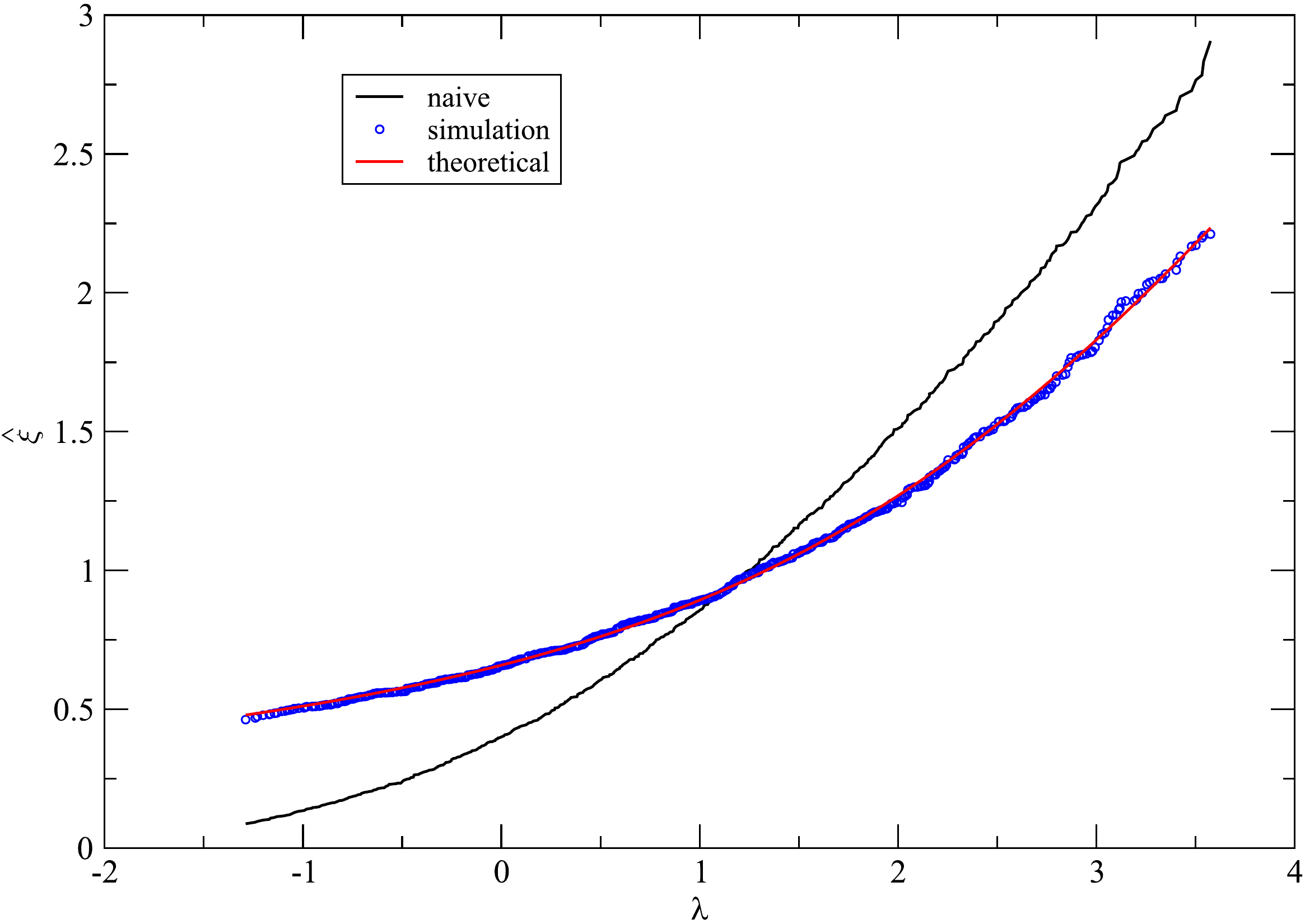} 
   \caption{Eigenvalues according to the optimal cleaning formula (\ref{oracle_gaussian_Wigner}) (red line) as a function of the observed noisy eigenvalues $\lambda$. The parameter are the same as in Fig. \ref{oracle_addition_density}. We also provide a comparison against the naive eigenvalues substitution method (black line) and we see that the optimal cleaning scheme indeed narrows the spacing between eigenvalues. }
   \label{oracle_addition}
\end{figure}

As a second example, we now consider a less trivial case and suppose that $\C$ is a 
white 
Wishart matrix with parameter $q_0$. For any $q_0 > 0$, it is well known that the Wishart matrix has nonnegative eigenvalues. However, we expect that the noisy effect coming from the GOE matrix pushes some true eigenvalues towards the negative side of the real axis. In Fig \ref{oracle_addition_density}, we clearly observe this effect and a good cleaning scheme should bring these negative eigenvalues back to positive values.
In order to use Eq. (\ref{oracle_gaussian}), we invoke once again the free addition formula to find the following equation for the Stieltjes transform of $\M$:
\begin{align*}
& 0 = -\; q_0\sigma^2\stj_{\M}(z)^3 + (\sigma^2 +q_0\,z)\stj_{\M}(z)^2  \\
& \qquad\qquad + (1-q_0-z) \stj_{\M}(z) + 1\,,
\end{align*}
for any $z = \lambda - \ii\eta$ with $\eta \rightarrow 0$. It then suffices to take the real part of the Stieltjes transform $\stj_{\M}(z)$ that solves this equation\footnote{We take the solution which has a strictly nonnegative imaginary part} to get the Hilbert transform. {In order to check formula Eq.\ \eqref{oracle_free_add} using numerical simulations, 
we have generated a matrix of $\M$ given by Eq.\ \eqref{free_additive_noise} with $\C$ a fixed white Wishart matrix with parameter $q_0$ and $\O\B \O^{\dag}$  a GOE matrix with radius 1. As we know exactly $\C$, we can compute numerically
the oracle estimator as given in \eqref{oracle_estimator} for each sample. In Fig.\ \ref{oracle_addition}, we see that our theoretical prediction in the large $N$ limit 
compares very nicely with the mean values of the empirical oracle estimator computed from the sample}. 
We can also notice in Fig. \ref{oracle_addition_density} that the spectrum of the cleaned matrix 
(represented by the ESD in blue) is narrower 
than the standard Mar{\v c}enko-Pastur density. This confirms the observation made in Sec. \ref{sec:RI_oracle_overlap}.

\subsection{Free multiplicative noise}
\label{sec:oracle_mult}

By proceeding in the same way as in the additive case, we can derive formally a nonlinear shrinkage estimator that depends on the observed eigenvalues $\lambda$ of $\M$ defined by
\begin{equation*}
\M = \sqrt{\C} \O \B \O^{\dag} \sqrt{\C}\,.
\end{equation*}
Following the computations done above, we can find after some manipulations of the global law estimate \eqref{global_law_multiplication}:
\begin{equation}
\label{oracle_MP_tmp}
\Tr \left({\G}_{\M}(z) \C\right) = N (z\stj_{\M}(z) - 1) \str_{\B}(z \stj_{\M}(z) - 1).
\end{equation}
Using the analyticity of the $\str$-transform, we define the function $\gamma_{\B}$ and $\omega_{\B}$ such that:
\begin{equation}
\label{anisotropic_s_transform_decomposition}
\underset{z \rightarrow \lambda - \ii0^{+}}{\lim}  \str_{\B}(z\stj_{\M}(z) - 1) := \gamma_{\B}(\lambda) + \ii\pi\rho_{\M}(\lambda) \omega_{\B}(\lambda)\,,
\end{equation}
and as a result, the optimal RIE (or nonlinear shrinkage formula) for the free multiplicative noise model (\ref{free_multiplicative_noise}) reads:
\begin{equation}
\label{oracle_elliptic_cov}
\widehat{\xi}_i = F_2(\lambda_i); \quad F_2(\lambda)= \lambda \gamma_{\B}(\lambda) + (\lambda \hil_{\M}(\lambda) - 1)\omega_{\B}(\lambda)\,,
\end{equation}
and this is the analog of the estimator \eqref{oracle_free_add} in the multiplicative case. 

\subsubsection{Empirical covariance matrix}

As a first application of the general result Eq. (\ref{oracle_elliptic_cov}), we reconsider the homogeneous Mar{\v c}enko-Pastur setting where $\B = \X \X^{\dag}$ with $\X$ defined as in Eq.\ \eqref{eq:empirical_covariance}. We trivially find from 
the definition of the $\str$-transform \eqref{eq:s_tranform_wishart} that (\ref{anisotropic_s_transform_decomposition}) yields in this case: 
\begin{eqnarray}
\gamma_{\B}(\lambda) & = & \frac{1-q+q\lambda \hil_{\M}(\lambda)}{|1-q+q\lambda \underset{z \rightarrow \lambda - \ii 0^{+}}{\lim} \stj_{\M}(z)|^2}\, \nonumber \\
 \omega_{\B}(\lambda) & = & - \frac{q\lambda}{|1-q+q\lambda \underset{z \rightarrow \lambda - \ii 0^{+}}{\lim} \stj_{\M}(z)|^2}\,. \nonumber \\
\end{eqnarray}
The nonlinear shrinkage function $F_2$ thus becomes:
\begin{equation}
\label{oracle_MP}
F_2(\lambda) = \frac{\lambda}{(1-q+q \lambda \hil_{\M}(\lambda))^2 + q^2 \lambda^2 \pi^2 \rho_{\M}^2(\lambda)},
\end{equation}
which is precisely the Ledoit-P\'ech\'e estimator derived in \cite{ledoit2011eigenvectors}. Let us insist once again on the fact that this is the oracle 
estimator, but it can be computed without the knowledge of $\C$ itself, but only with its noisy version $\M$. This ``miracle'' is of course only possible
thanks to the $N \to \infty$ limit that allows the spectral properties of $\M$ and $\C$ to become deterministically related one to the other.  

Like in the additive case, we can give a pretty insightful application of the formula (\ref{oracle_elliptic_cov}) based on Bayesian statistics once again. Let us suppose that $\C$ is a white inverse-Wishart matrix (\ie $\C^{-1}$ is a white Wishart matrix of parameter $q$). The eigenvalue distribution of $\C$ can then be computed exactly by performing the following change of variable\footnote{The factor $1-q$ is such that $\Tr \C = N$, which follows from the Mar{\v c}enko-Pastur equation.} $\mu = \left( (1-q) c \right )^{-1}$ in the Mar{\v c}enko-Pastur density function to get
\begin{eqnarray}
\label{densitys_IW}
\rho_{\C}(\mu) & = & \frac{\kappa}{\pi \mu^2} \sqrt{ (\mu_+ - \mu)(\mu - \mu_-)} \,,  \nonumber \\
\mu_{\pm} & \deq & \frac{1}{\kappa} [ \kappa + 1 \pm \sqrt{2\kappa + 1} ]\,,
\end{eqnarray}
with $q = (2\kappa + 1)^{-1}$ and $\kappa$ the hyper-parameter which is positive. From there, one can compute the corresponding Stieltjes transform of $\C$ 
\begin{equation}
\label{stieltjes_IW}
\stj_{\C}(z) = \frac{(1+\kappa)z - \kappa \pm  \kappa \sqrt{ (z-\mu_+)(z-\mu_-)} }{z^2},
\end{equation}
and we can also compute the Stieltjes transform of the perturbed matrix $\M$ thanks to the Mar{\v c}enko-Pastur equation \eqref{MP_equation}:
\begin{eqnarray}
\label{stieltjes_perturb_IW}
\stj_{\M}(z) & = & \frac{ z(1+\kappa) - \kappa(1-q) \pm \sqrt{ \psi(z,\kappa) }}{z(z+2q\kappa)}\,, \nonumber \\
\psi(z,\kappa) & \deq & (\kappa(1-q) - z(1+\kappa))^2 \nonumber \\
& & - z(z+2q\kappa)(2\kappa +1)
\end{eqnarray}
The reason why we insist on this matrix ensemble is that it plays a special role in multivariate statistics, especially for estimating covariance matrices because the famous linear shrinkage estimator \cite{haff1980empirical} turns out to be exact in this case, 
in the sense that it corresponds to the RIE as defined in the introduction. 
We can recover this result  within the present formalism. 
Indeed, the use of Eq. (\ref{stieltjes_perturb_IW}) in the estimator \eqref{oracle_MP} leads, after some computations, to:
\begin{equation}
\label{linear_shrinkage}
F_2({\lambda}) \;=\; \alpha \lambda + (1-\alpha), \quad \text{with } \alpha \;\deq\; \frac{1}{1+2q\kappa}\,.
\end{equation}
This is the linear shrinkage estimator that tells us to replace the noisy eigenvalues by a linear combination of the noisy eigenvalues and unity. 

\subsubsection{Elliptical Ensemble}
\label{sec:oracle_mult_elliptical}

\begin{figure}
   \includegraphics[scale = 0.32]{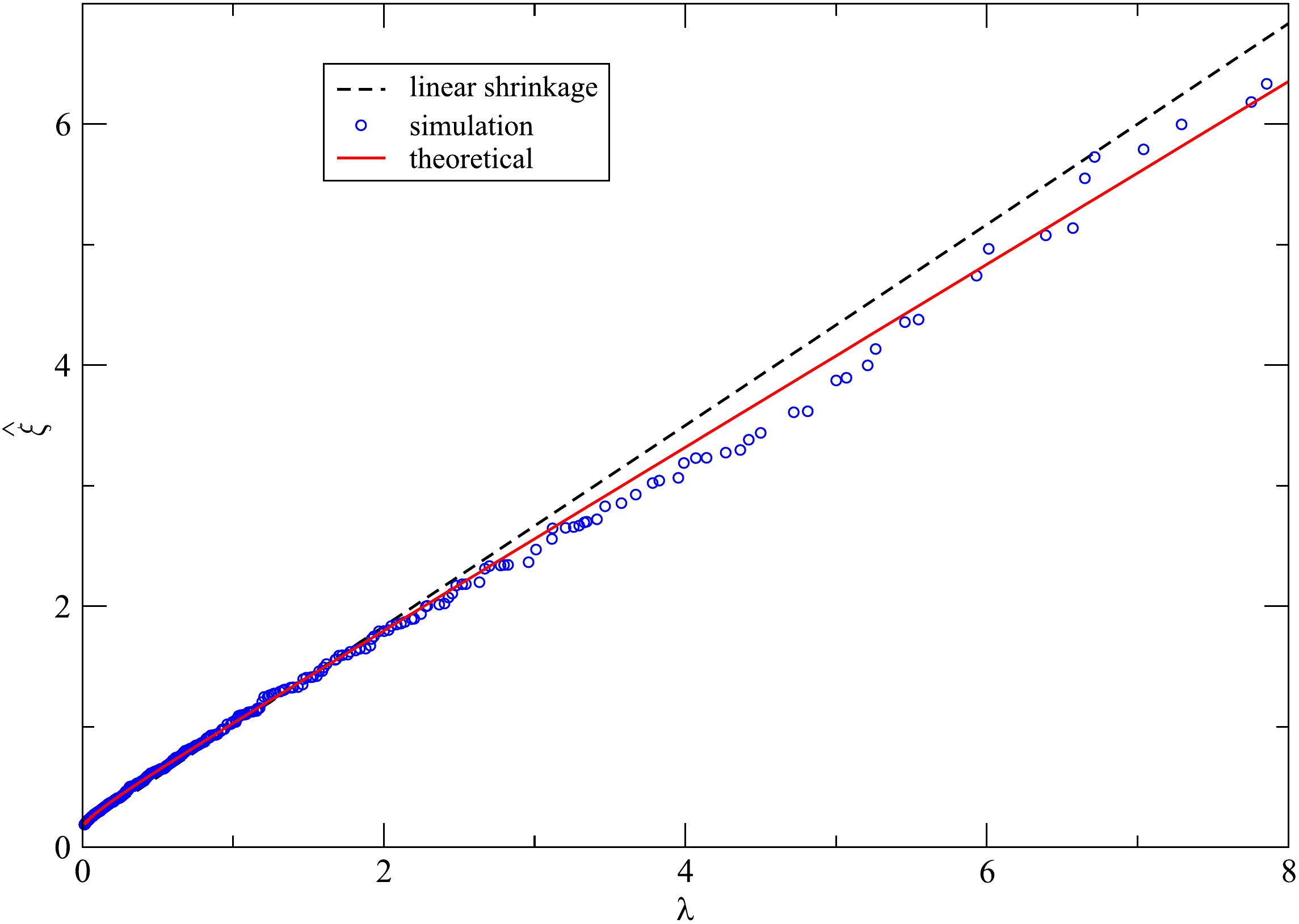} 
   \caption{Eigenvalues according to the optimal cleaning formula (\ref{oracle_elliptic_cov}) (red line) as a function of the observed noisy eigenvalues $\lambda$ when $\C$ is an inverse-Wishart matrix (with parameter $\kappa = 0.2$ and the $\Sigma = \text{diag}(\{\sigma_t^2\}_t)$ is distributed according the Mar{\v c}enko-Pastur density (with parameter $q_0 = 0.5$). We compare the result against numerical simulations (blue points) and the agreement is excellent. We furthermore compute the optimal cleaning scheme when $\Sigma = \It$ (black dotted line) and we see that $\Sigma$ allows one to go from linear to nonlinear shrinkage.}
   \label{oracle_cov}
\end{figure}

In this subsection, we now consider the elliptical model \eqref{covariance_elliptic} where the noise term is given by
\begin{equation*}
\B = \X \Sigma \X^{\dag},
\end{equation*}
for an arbitrary diagonal $T \times T$ matrix $\Sigma$. One immediately sees that the optimal shrinkage formula (\ref{oracle_elliptic_cov}) will now depends on $q = N/T$ and on the spectral measure of $\Sigma$, which prevents us to get a tractable form as in the homogeneous Mar{\v c}enko-Pastur case (\ref{oracle_MP}). However, we may expect to find a nonlinear shrinkage formula even when the signal is given by an Inverse-Wishart matrix. Indeed, the optimal cleaning formula is given by Eq.\ (\ref{oracle_elliptic_cov}) where we can compute numerically the $\str$-transform of $\B$ using the free multiplication and Eq. (\ref{S_dual_formula}) for any $\Sigma$. We illustrate this in Fig. (\ref{oracle_cov}) where the eigenvalues of $\Sigma$ are generated following Mar{\v c}enko-Pastur density and we see that the estimator \eqref{oracle_elliptic_cov} clearly deviates from the linear shrinkage \eqref{linear_shrinkage}. \\

As a second example, we consider the \textit{Student} ensemble of correlation matrices \cite{biroli2007student,el2009concentration} which has encountered some success in quantitative finance because it allows one 
to construct non-Gaussian correlated data with a clear interpretation of the matrix $\Sigma$. We impose the $\{\sigma_{t}^{2}\}_{t=1}^{T}$ to be distributed according  to an inverse-gamma distribution, \ie
\begin{equation}
\label{inverse_gamma}
\rho_{\Sigma}(\sigma^2) = \frac{1}{\Gamma\left(\frac{\mu}{2}\right)} \exp\left[-\frac{\sigma_0^2}{\sigma^2}\right] \frac{\sigma_0^{\mu}}{\sigma^{1+\mu}}\,,
\end{equation}
where we set $\sigma_0^2 := (\mu - 2)/2$ and $\mu > 2$ in order to have $\langle \sigma^2 \rangle = 1$. Within such prescription, the sample data $R_{it} := \sigma_t Y_{it}$, with $\langle Y_{it} Y_{jt'} \rangle \delta_{t,t'} = C_{ij}$, is characterized by the multivariate Student distribution of parameters $\mu$ and $N$ \cite{bouchaud2003theory}. From a financial perspective, this parametrization can be useful as a model where all individual stock returns are impacted by the same, time dependent scale factor $\sigma_t$ that represents the ``market volatility'' (see \cite{chicheportiche2012joint} for a discussion of this assumption). From empirical studies, one possible choice that matches quite well the data is to choose Eq. \eqref{inverse_gamma} with $\mu \approx 3 - 5$. The results above allow us to compute numerically either the LSD or the RIE for an arbitrary ``true'' signal $\C$, thus generalizing the work done in \cite{biroli2007student}.

\begin{figure}
   \includegraphics[scale = 0.32]{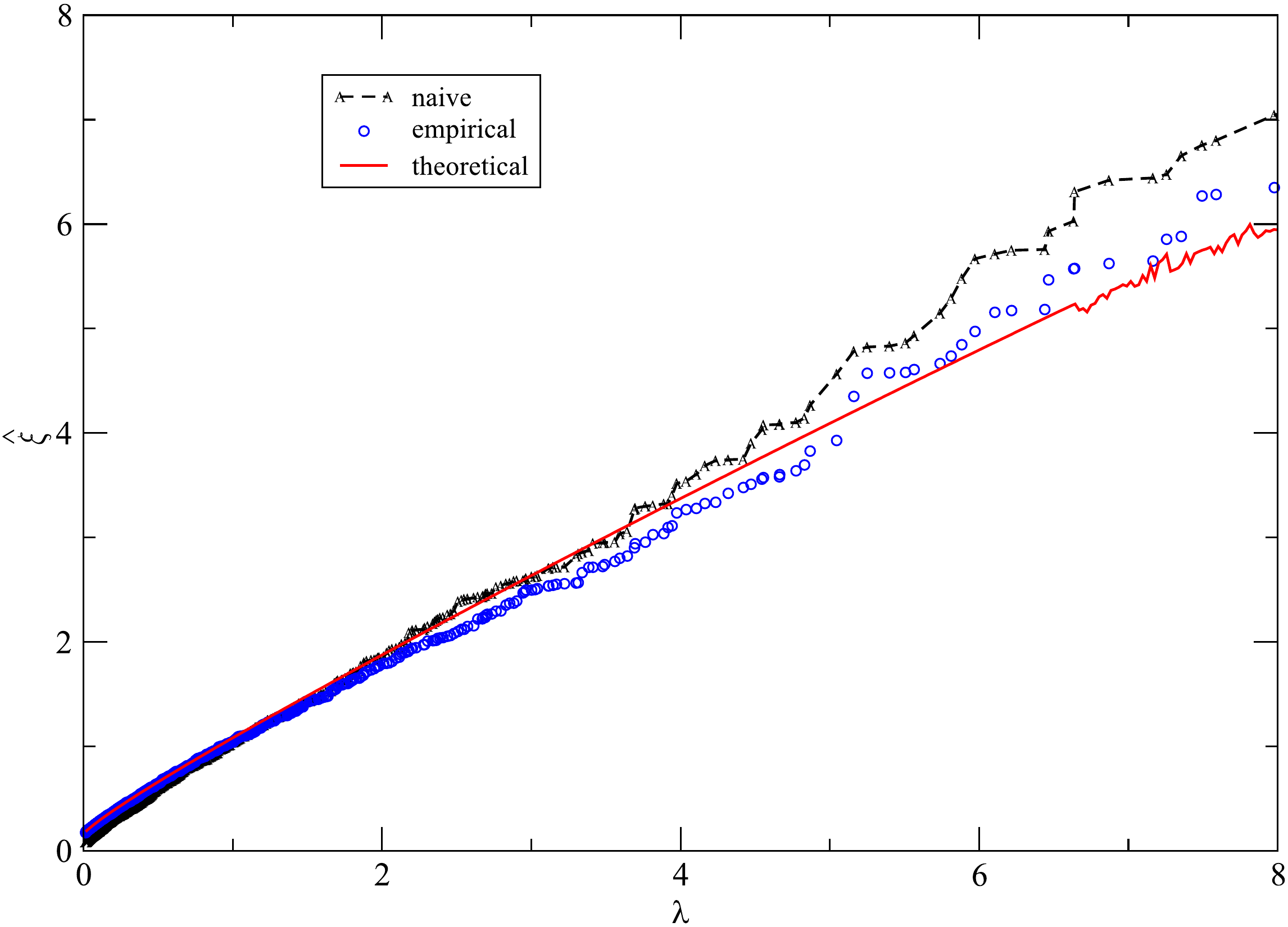} 
   \caption{Eigenvalues according to the optimal cleaning formula (\ref{oracle_elliptic_cov}) as a function of the noisy observed eigenvalues $\lambda$ when $\C$ is an inverse-Wishart matrix (with parameter $\kappa = 0.2$) and the $\Sigma = \text{diag}(\{\sigma_t^2\}_t)$ is generated according an inverse-gamma distribution \eqref{inverse_gamma} (with parameter $\mu = 6$). We compare the RIE (\ref{oracle_elliptic_cov}) (red line) against numerical simulations (blue points) and the agreement is quite convincing, especially in the bulk. We compare it with the substitution procedure (black dotted line) which leads to a wider spectrum.}
   \label{oracle_inverse_gamma}
\end{figure}

We plot in Fig. \ref{oracle_inverse_gamma} the RIE (\ref{oracle_elliptic_cov}) when the eigenvalues of $\Sigma$ are generated following the inverse-gamma distribution with $\mu = 6$ and $\C$ is still an inverse-Wishart matrix of parameter $\kappa = 0.2$. The numerical procedure is the same as for the previous example. The results we obtain are quite convincing, especially in the bulk. The noisy fluctuations for the largest eigenvalues in Fig. \ref{oracle_inverse_gamma} can be explained by the difficulty to solve Eq. (\ref{generalized_MP}) and \eqref{Z_cov_elliptic} outside of the bulk, most notably due to the inversion of the $\ttr$-transform of $\Sigma$. However, we see that these large eigenvalues still have the right behavior in the sense that they are shrunk downward compared to the ``naive'' substitution procedure.

\subsection{Information-Plus-Noise matrix}
\label{sec:oracle_info}

\begin{figure}
   \includegraphics[scale = 0.32]{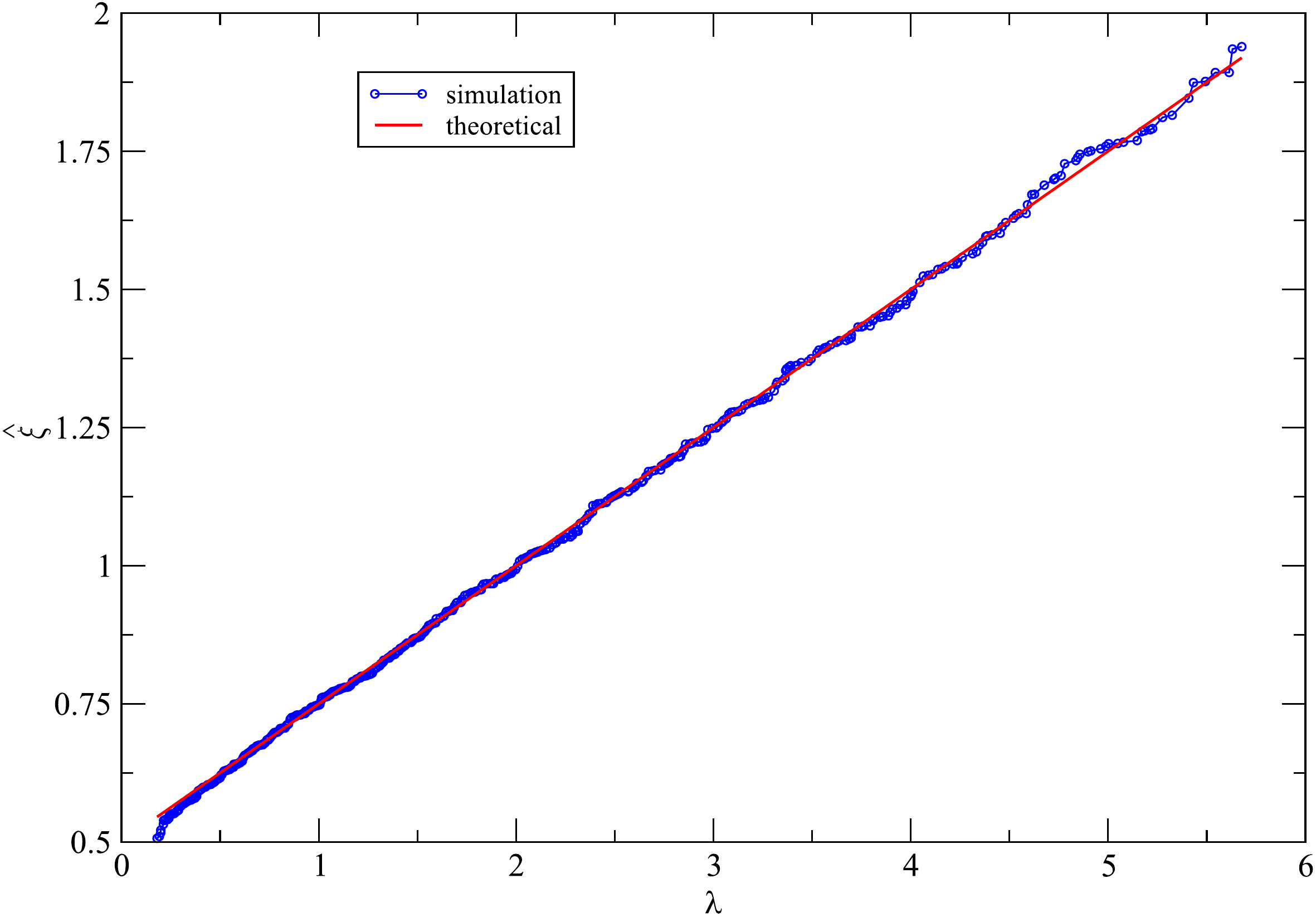} 
   \caption{Eigenvalues according to the optimal cleaning formula (\ref{info_noise_shrinkage}) as a function of the noisy observed eigenvalues $\lambda$ with the same setting than in Sec. (\ref{sec:overlaps_info}). We compare the estimator (\ref{info_noise_shrinkage}) (red line) against numerical simulations coming from a {\it single} sample with $N = 500$ (blue points), and the agreement is excellent. }
   \label{oracle_info}
\end{figure}

The derivation of the asymptotic RI estimator for the Information-Plus-Noise model is a bit more 
tedious compared to the previous cases but one can follow the same route to find the 
desired result. We leave the complete derivation to the reader; 
the final formula for the corresponding shrinkage function $F_3$ reads: 
\begin{equation}
\label{info_noise_shrinkage}
F_3({\lambda}) = \zeta(\lambda) (\lambda - \sigma^2(1-q) - 2q\sigma^2 \lambda \hil_{\M}(\lambda)) + q\sigma^2(1- \gamma_{\M}(\lambda)),
\end{equation}
where $\hil_{\M}(\lambda)$ is as before the Hilbert transform of the probability density $\rho_\M$, $\zeta(\lambda)$ is given in \eqref{overlap_info_noise} and the function $\gamma_{\M}$ is defined by
\begin{equation*}
\gamma_{\M}(\lambda) = \hil_{\M}(\lambda)(\lambda - \sigma^2(1-q)) + q\sigma^2\lambda (\pi^2 \rho_{\M}^2(\lambda) - \hil_{\M}^2(\lambda)).
\end{equation*}
If we consider the trivial case of zero noise (\ie $\sigma = 0$), we have by definition that $\M = \C$ and we indeed see this in Eq. (\ref{info_noise_shrinkage}) where the optimal shrinkage formula becomes $\widehat{\xi}_i  = \lambda_i$. The other limit that can be studied without much effort is when the sample size becomes much larger than the number of variable (\ie $q = 0$). In this case, we know that $\M = \C + \sigma^2 \In$ by the law of large number. The optimal shrinkage \eqref{info_noise_shrinkage} 
gives in that case $\widehat{\xi}_i = \lambda_i - \sigma^2$ which was expected because the observation matrix $\M$ is simply a shift of the signal by a factor $\sigma^2$. Let us now reconsider the same numerical example of Section (\ref{sec:overlaps_info}) and we apply the same procedure to test the RIE (\ref{info_noise_shrinkage}) than the last two sections. We clearly see in Fig. \ref{oracle_info} that the agreement is remarkable even for a finite $N$.

\section{Conclusion and open problems}

As we recalled in the introduction, RMT is already at the heart of many significant contributions when it comes to reconstructing a true signal matrix $\C$ of large dimension from a noisy measurement. In this paper, we revisited this statistical problem and considered the so-called oracle estimator which is optimal with respect to the Euclidean norm. In particular, we have established the global resolvent law for three distinct ensembles of random matrices that embrace well-known models like the deformed Wigner or the sample covariance matrix.  These results on the asymptotic convergence have two important applications: (i) they allow us to find 
exact results on the overlap between the eigenvectors of the signal matrix with the corrupted ones; (ii) most importantly, they lead to the `miracle' 
which allows the oracle estimator to be expressed {\it without any knowledge} of the signal $\C$ in the large $N$ limit. This last observation, that generalizes the work of Ledoit and P\'ech\'e~\cite{ledoit2011eigenvectors}, 
should be of particular interest in practical cases. 

We emphasize that the proposed estimators are \textit{optimal} (in the $\mathbb{L}^2$ norm sense) when the dimension of the problem becomes very large and under no particular prior beliefs on the eigenvector structure of the true matrix $\C$.    
However, it happens in practice that one could have a prior structure on the eigenvectors of $\C$ (factor models), 
and it would be interesting to see how can we rewrite our problem in a non-RI framework. 
This natural extension is left for future work. Note that the case of so-called `spiked' eigenvalues/eigenvectors in the context of empirical covariance matrix was recently solved in \cite{bun2016cleaning}.

We also highlighted the interesting connexion between our work with some famous result of Bayesian statistics. 
For instance, we found out that our results generalize the Wiener filter \cite{wiener1949extrapolation} 
(additive case) but also the linear shrinkage \cite{haff1980empirical, ledoit2004well} (multiplicative case), 
and both have encountered many successes in practical cases. Moreover, the Bayesian theory 
has found several applications in modern statistical analysis, especially because the 
large amount of data may allow one to identify a pattern in the data which 
could be used as a prior. The different estimators we proposed are powerful in the sense that 
they depend only on {\it observable} quantities.  Differently said, they can be used in many different contexts with
no parameter to fit. This is in adequation with the recent work of \cite{ledoit2016numerical} and \cite{bun2016cleaning} which provide efficient numerical methods to use 
these (limiting) results for real life applications, in the specific case of sample covariance matrices. One could also extend the above results to more realistic models of covariances that accounts 
for autocorrelation effects \cite{burda2005spectral} or fat-tailed distributions \cite{couillet2015random}. 


Although our computations are based on the \textit{non-rigorous} Replica method, the comparison between our theoretical formulas and empirical simulations demonstrates the robustness of each proposed estimator. Hence, one can certainly think of possible extensions of this work  based on the same method. For instance, a natural extension for our free additive perturbation model would be given by
\begin{equation}
\M = \C + O_q {\bf B} O_q^{\dag}
\end{equation}
where the law of the matrix $O_q\in O(N)$ interpolates between the Haar measure on the 
Orthogonal group $O(N)$ when $q = 0$ and a given measure on the permutation group when 
$q = +\infty$. Differently said, $\M$ interpolates between the free and the classical addition. A natural prescription would be to imagine that $O_q$ is the result of Biane's Brownian motion \cite{biane1997free} on $O(N)$. Another natural possibility is to assume that $O_q$ is distributed according to the probability measure with the Harish-Chandra-Itzykson-Zuber (HCIZ) weight \cite{harish1957differential}, \cite{itzykson1980planar}:
\begin{equation}
{\cal P}_q(dO) \propto \exp\left[ q N \Tr \C O {\bf B} O^{\dag} \right]dO
\end{equation}
which has the right limits when $q \to 0$ (Haar measure on the orthogonal group) and $q \to \infty$ (deterministic measure 
rearranging the spectrum of $B$ in non-increasing order). Hence, considering a replica method for this specific case might give us access to the global law of the resolvent of $\M$ which should enable us to express the correlation function of angular integrals, leading to an alternative expression of the Morosov-Shatashvili formula \cite{morozov1992pair,shatashvili1993correlation} expressed in terms of the free energy of HCIZ integrals.

\section*{Acknowledgments}

We are indebted to M. Abeille and S. P{\'e}ch{\'e} for fruitful discussions. RA received funding from the European Research Council under the European Union's Seventh Framework Programme (FP7/2007-2013)/ERC grant agreement nr. 258237 and thanks the Statslab in DPMMS, Cambridge for its hospitality.

\bibliographystyle{IEEEtran}
\bibliography{IEEEabrv,RIE_noisy}

\appendices 
\section{Reminder on transforms in RMT}
\label{sec:appendix_transform}

We give in this first appendix a short reminder on the different transforms that are useful in the study of the statistics of eigenvalues in RMT due to their link with free probability theory (see \eg \cite{speicher2009free} or \cite{burda2013free} for a review). The resolvent of $\M$ is defined by:
\begin{equation}
\label{resolvent}
{\G}_{\M}(z) \;\deq\; (z\In - \M)^{-1},
\end{equation}
and the \textit{Stieltjes} (or sometimes {\it Cauchy}) transform is the normalized trace of the Resolvent:
\begin{eqnarray}
\label{app_stieltjes}
\stj_{\M}(z) & \;\deq\; & \frac1N \Tr {\G}_{\M}(z)\,, \nonumber \\
& = & \frac1N \sum_{k=1}^N \frac{1}{z - \lambda_k} \underset{N \rightarrow \infty}{\sim} \int \frac{\rho_{\M}(\lambda)}{z- \lambda} \dd\lambda\,.\nonumber\\
\end{eqnarray}
The Stieltjes transform can be interpreted as the average law and is very convenient in order to describe the convergence of the eigenvalues density $\rho_{\M}$. If we set $z = \lambda - \ii\eta$ and take the limit $\eta \rightarrow 0$, we have in the large $N$ limit
\begin{equation*}
G_{\M}(\lambda_i - \ii\eta) = \text{P.V.} \int \frac{\rho_{\M}(\lambda')}{\lambda - \lambda'} \dd\lambda' + \ii \pi \rho_{\M}(\lambda)
\end{equation*}
where the real part is often called the \textit{Hilbert} transform $\hil_{\M}(\lambda)$ and the imaginary part leads to the eigenvalues density. 

When we consider the case of adding two random matrices that are (asymptotically) free with each other, it is suitable to introduce the functional inverse of the Stieltjes transform known as the {\it Blue} transform \cite{zee1996law}
\begin{equation}
\label{app_blue}
\btr_{\M}(\stj_{\M}(z)) = z.
\end{equation}
This allows us to define the so-called $\rtr$-transform
\begin{equation}
\label{app_red}
\rtr_{\M}(z) \;\deq\; \btr_{\M}(z) - \frac1z,
\end{equation}
which can be seen as the analogue in RMT of the logarithm of the Fourier transform for free additive convolution. More precisely, if $\AAA$ and $\B$ are two $N \times N$ independent invariant symmetric random matrices, then in the large $N$ limit, the spectral measure of $\M = \AAA + \B$ is given by 
\begin{equation}
\label{app_free_add}
\rtr_{\M}(z) = \rtr_{\AAA}(z) + \rtr_{\B}(z),
\end{equation}
known as the free addition formula \cite{voiculescu1992free}. In this case, we note by $\rho_{\AAA \bxp \B}$ the eigenvalues density of $\M$. 

We can do the same for the free multiplicative convolution. In this case, we rather have to define the so-called $\ttr$ (or sometimes $\eta$ \cite{tulino2004random} ) transform given by
\begin{equation}
\label{app_T}
\ttr_{\M}(z) = \int \frac{\rho_{\M}(\lambda) \lambda}{z- \lambda} \dd\lambda  \equiv z \stj_{\M}(z) - 1,
\end{equation}
which can be seen as the moment generating function of $\M$. Then, the $\str$-transform of $\M$ is then defined as
\begin{equation}
\label{app_S}
\str_{\M}(z) := \frac{z+1}{z \ttr^{-1}_{\M}(z)} 
\end{equation}
where $\ttr^{-1}_{\M}(z)$ is the functional inverse of the $\ttr$-transform. Before showing why the $\str$-transform is important in RMT, one has to be careful about the notion of product of free matrices. Indeed, if we reconsider the two $N \times N$ independent symmetric random matrices $\AAA$ and $\B$, the product $\AAA\B$ is in general not self-adjoint even if $\AAA$ and $\B$ are self-adjoint. However, if $\AAA$ is positive definite, then the product $\sqrt{\AAA} \B \sqrt{\AAA}$ makes sense and share the same moments than the product $\AAA\B$. We can thus study the spectral measure of $\M = \sqrt{\AAA}\B\sqrt{\AAA}$ in order to get the distribution of the free multiplicative convolution $\rho_{\AAA \bxt \B}$. The result, first obtained in \cite{voiculescu1992free}, reads: 
\begin{equation}
\str_{\AAA \bxt \B}(z) := \str_{\M}(z) = \str_{\AAA}(z) \str_{\B}(z).
\end{equation}
The $\str$-transform is therefore the analogue of the Fourier transform for free multiplicative convolution. 

\section{Derivation of the global law estimate}\label{appendixB}

\subsection{The replica method}
\label{sec:appendix_replica}

 Throughout the following, we shall use the abbreviation $\G(z) \equiv \G_\M(z)$ for simplicity.  The starting point of our approach is to rewrite the entries of the resolvent $\G(z) \deq (G_{ij}(z)) \in \mathbb{R}^{N\times N}$ 
 using the Gaussian integral representation of an inverse matrix:
\begin{align}
\label{resolvent_entries}
& G_{ij}(z) = \nonumber\\
& \frac{\int \eta_i \eta_j \exp\left\{\frac12 \sum_{k,l=1}^{N} \eta_k (z\delta_{k,l} - \M_{kl}) \eta_l \right\} \prod_{k=1}^{N} \dd\eta_k}{\int \exp\left\{-\frac12 \sum_{k,l=1}^{N} \eta_k (z\delta_{kl} - \M_{kl}) \eta_l \right\} \prod_{k=1}^{N} \dd\eta_k}.
\end{align}
We recall that the claim is that for a complex $z$ not too close to the real axis, we expect the resolvent to be self-averaging in the large $N$ limit, that is to say independent of the specific realization of the matrix itself. Therefore we can study the resolvent ${\G}_{\M}(z)$ through its ensemble average (denoted by $\langle \cdot \rangle$ in the following) given by:
\small
\begin{align}
\label{avg_resolvent_entries}
&\left\langle G_{ij}(z) \right\rangle = \nonumber  \\ 
& \left\langle \frac{1}{{\cal Z}} \int \eta_i \eta_j \exp\left[-\frac12 \sum_{k,l=1}^{N} \eta_k (z\delta_{kl} - M_{kl}) \eta_l \right] \prod_{k=1}^{N} \dd\eta_k  \right\rangle
\end{align}
\normalsize
where ${\cal Z}$ is the partition function, i.e. the denominator in Eq. (\ref{resolvent_entries}). The computation of the average value is highly non trivial in the general case. The replica method tells us that the expectation value can be handled thanks to the following identity
\footnotesize
\begin{eqnarray}
\label{eq:replica_trick}
\left\langle G_{ij}(z) \right\rangle & = & \underset{n \rightarrow 0}{\lim} \avgBB{{\cal Z}^{n-1} \int \pBB{\prod_{k=1}^{N} \dd\eta_k  }\eta_i \eta_j \times  \nonumber \\
& &  \exp\left[-\frac12 \sum_{k,l=1}^{N} \eta_k (z\delta_{k,l} - M_{kl}) \eta_l \right]  }\nonumber \\
& = & \underset{n \rightarrow 0}{\lim} \int \left(\prod_{k=1}^{N} \prod_{\alpha=1}^{n} d\eta_k^{\alpha}\right) \eta^{1}_i \eta^{1}_j \times \nonumber \\
& & \left\langle \exp\left\{-\frac12 \sum_{\alpha=1}^{n} \sum_{k,l=1}^{N} \eta_k^{\alpha} (z\delta_{k,l} - \M_{kl}) \eta_l^{\alpha} \right\} \right\rangle. \nonumber\\
\end{eqnarray}
\normalsize
We have thus transformed our problem to the computation of $n$ replicas of the initial system (\ref{resolvent_entries}). So when we have computed the average value in \eqref{avg_resolvent_entries}, it suffices to perform an analytical continuation of the result to real values of $n$ and finally takes the limit $n \rightarrow 0$. The main concern of this non-rigorous approach is that we assume that the analytical continuation can be done with only $n$ different set of points which could lead to uncontrolled approximation in some cases \cite{parisi1980sequence}.  We stress that the replica trick \eqref{eq:replica_trick} allows to investigate each entry $i,j \in \{1,\dots,N\}$ of the resolvent $\G_\M$. This result we get is thus stronger than classical tools from free probability theory which concerns only the Stieltjes transform, i.e.\ the normalized trace of the resolvent. 

\subsection{Free additive noise}
\label{sec:appendix_add}

We consider a model of the form
\begin{equation*}
\M \;=\; \C + \O {\bf B} \O^{\dag}
\end{equation*}
where ${\bf B}$ is a fixed matrix with eigenvalues $b_1 > b_2 > \dots > b_N$ with spectral $\rho_{\B}$ and $\O$ is a random matrix chosen in the Orthogonal group $O(N)$ according to the Haar measure. {Clearly, the noise term is invariant under rotation so that we expect the resolvent of $\M$ to be in the same basis than $\C$. We therefore set without loss of generality that $\C$ is diagonal.}  Throughout the following, we fix $i,j \in \{1,\dots,N\}$ with possibly $j \neq i$.  In order to derive the global law estimate for the resolvent of the matrix $\M$, we have to consider the ensemble average value of the resolvent over the Haar measure for the $O(N)$ group, which can be written as follow
\small
\begin{align}
\label{free_convolution_orthogonal_avg}
&\left\langle G_{ij}(z) \right\rangle = \int \left(\prod_{\alpha = 1}^{n} \prod_{k=1}^{N} d\eta_k^{\alpha}\right) \eta_i^1 \eta_j^1  \prod_{\alpha = 1}^{n} e^{-\frac{1}{2} \sum_{k=1}^{N} (\eta_k^{\alpha})^2 (z - c_k) }  \nonumber \\
& \qquad\qquad\quad  \times \left\langle e^{-\frac{1}{2} \sum_{k,l=1}^{N} \eta_k^{\alpha} (\O\B \O^{\dag})_{kl} \eta_l^{\alpha}} \right\rangle_{\O}.
\end{align}
\normalsize
The evaluation of the later equation can be done straightforwardly if we set the measure $\dd\O$ to be a flat measure constrained to the fact that $\O \O^{\dag} = \In$, or equivalently said:
\small
\begin{equation*}
{\cal D}\O \propto \prod_{i,j=1}^{N} \dd O_{ij} \prod_{i,j=1}^{N} \delta\left( \sum_{k=1}^N O_{ik}O_{jk} - \delta_{i,j} \right)
\end{equation*}
\normalsize
where $\delta(\cdot)$ is the Dirac delta function and $\delta_{i,j}$ is Kronecker delta. In the case where $n$ is finite (and independent of $N$), one can notice that Eq. (\ref{free_convolution_orthogonal_avg}) is the Orthogonal low-rank version of the Harish-Chandra-Itzykson-Zuber integrals (\cite{harish1957differential}, \cite{itzykson1980planar}). The result is known for all symmetry groups (\cite{marinari1994replica, tanaka2008asymptotics} or \cite{guionnet2004asymptotic} for a more rigorous derivation), and reads for the rank-$n$ case
\small
\begin{align}
\label{free_convolution_HCIZ_rank_one}
& \int {\cal D}\O \exp\left[ \Tr \left( \frac12 \sum_{\alpha=1}^{n} \eta^{\alpha}(\eta^{\alpha})^{\dag} \O \B \O^{\dag}\right) \right] \\
& \qquad\qquad =\; \exp\left[ {\frac{N}{2} \sum_{\alpha = 1}^{n} \wtr_{\B}\left(\frac1N(\eta^{\alpha})^{\dag}\eta^{\alpha}\right)} \right], 
\end{align}
\normalsize
with $\wtr_B$ the primitive of the $\rtr$-transform of $B$.  We emphasize that \eqref{free_convolution_HCIZ_rank_one} can be obtained using a simple saddle-point calculation valid when $n$ is finite, in the limit $N \to \infty$ and its explicit structure simplifies a lot the following computations compared to standard approaches that uses Weingarten calculus (see e.g.\ \cite{collins2006integration}). 

By plugging \eqref{free_convolution_HCIZ_rank_one} into \eqref{free_convolution_orthogonal_avg}, we see that we need to compute:
\small
\begin{align*}
& \left\langle G_{ij}(z) \right\rangle = \int \left(\prod_{k=1}^{N} d\eta_k\right) \eta^{1}_i \eta^{1}_j  \times \\
& \exp\left\{ \frac{N}{2}\sum_{\alpha=1}^{n} \left[ \wtr_{\B}\left(\frac{(\eta^{\alpha})^{\dag}\eta^{\alpha}}{N} \right) -\frac{1}{2} \sum_{k=1}^{N} (\eta_k^{\alpha})^2 (z - c_k) \right] \right\},
\end{align*}
\normalsize
where we introduced a Lagrange multiplier $p^{\alpha} = \frac1N (\eta^{\alpha})^{\dag}\eta^{\alpha}$ which gives using Fourier transform (renaming $\zeta^{\alpha} = - 2i\zeta^{\alpha}/N$)
\footnotesize
\begin{align}
& \left\langle G_{ij}(z) \right\rangle \propto \int \int \int \left( \prod_{\alpha = 1}^{n} \dd p^{\alpha} \right) \left(\prod_{\alpha=1}^{n} \dd\zeta^{\alpha} \right) \left(\prod_{\alpha = 1}^{n}\prod_{k=1}^{N} d\eta_k^{\alpha} \right) \times  \nonumber \\
& \qquad\qquad \exp\left\{\frac{N}{2} \sum_{\alpha = 1}^{n} \left[ \wtr_{\B}(p^{\alpha}) - p^{\alpha}\zeta^{\alpha} \right] \right\} \times \nonumber \\
& \qquad\qquad \eta^{1}_i \eta^{1}_j  \exp\left\{ -\frac{1}{2} \sum_{\alpha = 1}^{n} \sum_{k=1}^{N} (\eta_k^{\alpha})^2 (z - \zeta^{\alpha} - c_k) \right\} \nonumber.
\end{align}
\normalsize
This additional constraint allows to retrieve a Gaussian integral over the $\{\eta_j\}$ which can be computed exactly. Ignoring normalization terms, we obtain
\small
\begin{align}
\label{eq:global_law_tmp}
& \left\langle G_{ij}(z) \right\rangle \propto \int \int  \frac{\delta_{i,j}}{z + \zeta^{1} - c_i} \times \nonumber \\
& \qquad\quad\exp\left\{ -\frac{Nn}{2} F_0(p^{\alpha}, \zeta^{\alpha}) \right\} \left( \prod_{\alpha=1}^{n}  \dd p^{\alpha} \dd\zeta^{\alpha}\right) 
\end{align}
\normalsize
where the `free energy' $F_0$ is given by
\small
\begin{eqnarray}
\label{free_convolution_energy}
F_0(p, \zeta) & \deq & \frac{1}{Nn} \sum_{\alpha = 1}^{n} \left[ \; \sum_{k=1}^{N} \log(z - \zeta^{\alpha} - c_k) \right. \nonumber \\ 
& &  - \; \wtr_{\B}(p^{\alpha}) + p^{\alpha}\zeta^{\alpha} \Biggl].
\end{eqnarray}
In the large $N$ limit, the integral can be evaluated by considering the saddle-point of the free energy $F_0$ as the other term is obviously sub-leading. We now use the \textit{replica symmetric} ansatz that tells us if the free energy is invariant under the action of the symmetry group $O(N)$, then we expect a saddle-point which is also invariant. This implies that we have at the saddle-point
\begin{equation}
\label{replica_symmetry}
p^{\alpha} = p \quad \text{ and } \quad \zeta^{\alpha} = \zeta\,, \qquad \forall \alpha \in \{1, \dots, n\},
\end{equation}
from which we find the following solution:
\begin{equation*}
\begin{dcases}
\zeta^* = \rtr_{\B}(p^*) \\
p^* = \stj_{\C}(z - \zeta^*)\,.
\end{dcases}
\end{equation*}
The trick is to see that we can get rid of one variable by taking the normalized trace of the (average) resolvent \eqref{eq:global_law_tmp} which gives the following relation for the Stieltjes transform: $ \stj_{\M}(z) = \stj_{\C}(z - \rtr_{\B}(p^*)) = p^*$, which is equivalent to:
\begin{equation}
\label{free_addition_stieltjes}
p^{*} = \stj_{\M}(z) =  \stj_{\C}\left(z - \rtr_{\B}\left(\stj_{\M}(z) \right)\right),
\end{equation}
and therefore
\begin{equation}
\label{eq:zeta_star}
	\zeta^* = \rtr_{\B}(\stj_{\M}(z)).
\end{equation}

In conclusion, by plugging \eqref{free_addition_stieltjes} and \eqref{eq:zeta_star} into \eqref{eq:global_law_tmp} and then taking the limit $n \rightarrow 0$, we obtain the global law estimate
\begin{equation}
\left\langle G_{ij}(z) \right\rangle = (Z(z)\In - \C)^{-1}_{i,i} \delta_{i,j} 
\end{equation}
with
\begin{equation}
Z(z) = z - \rtr_{\B} \left(\stj_{\M}(z)\right),
\end{equation}
which gives exactly the  matrix  result stated in Eq.\ (\ref{global_law_addition}) and (\ref{Z_addition})  since the indexes $i$ and $j$ are arbitrary.

\subsection{Free multiplicative noise}
\label{sec:appendix_mult}

Let us set the measurement matrix $\M$ as:
\begin{equation}
\M = \sqrt{\C} \O \B \O^{\dag} \sqrt{\C}
\end{equation}
where $\O$ is still a rotation matrix over the Orthogonal group, $\C$ is a positive definite matrix and $\B$ is such that $\Tr \B \neq 0$. {Note that we can assume without loss of generality that $\C$ is diagonal because the argument of the previous subsection still applies.}  As in the previous section, we may fix $i,j \in \{1,\dots,N\}$ throughout the following without loss of generality as the generalization to the matrix relation \eqref{global_law_multiplication} will be immediate.  Using the abbreviation $\G(z) \equiv \G_\M(z)$, the replica identity \eqref{eq:replica_trick} allows us to write the entries of the resolvent of $\M$ as follow
\begin{align}
& \left\langle G_{ij}(z) \right\rangle = \int \left(\prod_{\alpha = 1}^{n} \prod_{k=1}^{N} d\eta_k^{\alpha}\right) \eta_i^{1} \eta_j^{1} e^{-\frac{z}{2} \sum_{\alpha = 1}^{n} \sum_{k=1}^{N} (\eta_k^{\alpha})^2 } \times \nonumber\\
& \qquad\qquad \left\langle e^{ \frac{1}{2} \sum_{k,l=1}^{N} \sum_{\alpha = 1}^{n} \eta_k^{\alpha} (\sqrt{\C} \O\B \O^{\dag}\sqrt{\C} )_{kl} \eta_l^{\alpha} } \right\rangle_{\O}.
\end{align}
We can notice that the matrix $\sum_{\alpha = 1}^{n} \left(\sqrt{\C}\,  \eta^{\alpha}\right) \left(\sqrt{\C}\,  \eta^{\alpha}\right)^{\dag}$
is a symmetric rank-$n$ matrix, with $n$ finite and independent of $N$. Therefore, the expectation over the Haar measure still leads to a rank-$n$ Orthogonal version of HCIZ integral, and the result reads \cite{guionnet2004asymptotic}:
\begin{align}
& \left\langle \exp\left\{ \frac{1}{2} \sum_{k,l=1}^{N} \sum_{\alpha = 1}^{n} \eta_k^{\alpha} (\sqrt{\C}  \O\B \O^{\dag} \sqrt{\C} )_{kl} \eta_l^{\alpha} \right\} \right\rangle_{\O} \nonumber\\
& \qquad\quad =\; \exp\left[\frac{N}{2} \sum_{\alpha=1}^{n} \wtr_{\B}\left( \frac1N  \sum_{i=1}^{N} (\eta_i^{\alpha})^2 c_i \right) \right].
\end{align}
As in the free addition case, let us defined the auxiliary variable $p^{\alpha} =\frac1N \sum_{i=1}^{N} (\eta_i^{\alpha})^2 c_i $ that we enforce by a Dirac delta function. By following the same step as in the previous section, we get a Gaussian integral over the $\{\eta_k^{\alpha}\}$ that we compute to eventually find that
\begin{align}
& \left\langle G_{ij}(z) \right\rangle \propto \int \int \frac{\delta_{i,j}}{z - \zeta^1 c_i} \times \nonumber \\
& \qquad\qquad \exp\left\{- \frac{Nn}{2} F_{0}(p^{\alpha}, \zeta^{\alpha}) \right\} \prod_{\alpha = 1}^{n} \dd p^{\alpha} d\zeta^{\alpha}
\end{align}
where the free energy is given by
\begin{align}
& F_{0}(p^{\alpha}, \zeta^{\alpha}) \;\deq\; \frac{1}{n} \sum_{\alpha=1}^{n} \Biggl[ \; \frac{1}{N} \sum_{k=1}^{N} \log(z - \zeta^{\alpha} c_k ) \nonumber \\
& \qquad\qquad\qquad + \zeta^{\alpha} p^{\alpha} - \wtr_{\B}(p^{\alpha}) \Biggl].
\end{align}
We now assume that the saddle-point solution can be computed using the replica symmetry ansatz \eqref{replica_symmetry}, so that the free energy becomes 
\begin{equation}
F_{0}(p^{\alpha}, \zeta^{\alpha}) \equiv F_{0}(p, \zeta) = \frac{1}{N} \sum_{k=1}^{N} \log(z - \zeta c_k ) + \zeta p - \wtr_{\B}(p).
\end{equation}
We first consider the derivative w.r.t. p which leads to
\begin{equation}
\label{eq:zeta_star_mult}
\zeta^{*} = \rtr_{\B}(p).
\end{equation}
The other derivative gives 
\begin{equation}
\label{eq:p_star_mult_tmp}
p^{*} = \frac{\ttr_{\C} \left( \frac{z}{\rtr_{\B}(p^*)} \right)}{\rtr_{\B}(p^*)}.
\end{equation}
Hence, we see that the resolvent is given in the large $N$ limit and the limit $n \rightarrow 0$ by 
\begin{equation}
\left\langle G_{ij}(z) \right\rangle = \frac{\delta_{i,j}}{z - \rtr_{\B}(p^*) c_i}.
\end{equation}
It now remains to determine $p^*$ and to that end, we can find a genuine simplification in this last expression by using the connexion with the free multiplication convolution. By taking the normalized trace in the latter equation, it yields
\begin{equation}
\label{eq:global_law_mult_tmp}
z \stj_{\M}(z) = Z \stj_{\C}(Z), \quad\text{with }\quad Z = \frac{z}{\rtr_{\B}(p^*)},
\end{equation}
which can be easily rewrite using \eqref{app_T} as 
\begin{equation*}
\ttr_{\M}(z) = \ttr_{\C}(Z).
\end{equation*}
Next, let us define 
\begin{equation}
	\label{eq:x_T}
	\omega = \ttr_{\M}(z) = \ttr_{\C}(Z)\,
\end{equation} 
which implies from \eqref{eq:p_star_mult_tmp} that $p^* = \omega/\rtr_{\B}(p^*)$. Then, we may rewrite the Eq.\ \eqref{eq:x_T} as 
\begin{equation*}
\label{tmp}
z \ttr_{\M}(z) = Z \ttr_{\C}(Z) \rtr_{\B}(p^*)\,.
\end{equation*}
It is trivial to see that plugging $\omega$ into this latter expression yields $\omega \ttr_{\M}^{-1}(\omega) = \omega \ttr_{\C}^{-1}(\omega) \rtr_{\B}(p^*)$, and by the definition \eqref{app_S} of the $\str$-transform, we have 
\begin{equation}
\label{eq:S_transform_tmp}
\str_{\M}(\omega) = \frac{\str_{\C}(\omega) }{\rtr_{\B}(p^*)}.
\end{equation}
In order to retrieve the desired result, we use the following relation 
\begin{equation}
\label{R_S_transform}
\frac{1}{\rtr_{\B}(p^*)} = \str_{\B}(p^* \rtr_{\B}(p^*)),
\end{equation}
that follows from the very definition of the $\str$-transform of $\B$. But recalling that $p^* = \omega/\rtr_{\B}(p^*)$, we conclude from \eqref{eq:zeta_star_mult}, \eqref{eq:x_T} and \eqref{R_S_transform} that
\begin{equation}
	\label{eq:zeta_star_mult_sol}
	\frac{1}{\zeta^{*}} = \rtr_{\B}(p^*) = \str_{\B}(\ttr_{\M}(z)).
\end{equation}
Then going back to \eqref{eq:S_transform_tmp}, we see that the spectral density of $\M$ is given by the free multiplication 
\begin{equation}
\str_{\M}(\omega) = \str_{\C}(\omega) \str_{\B}(\omega),
\end{equation}
as expected and it proves that the replica symmetry ansatz holds in this model. Finally, by plugging \eqref{eq:zeta_star_mult_sol} into \eqref{eq:global_law_mult_tmp}, we conclude that we can characterize the asymptotic global law of the resolvent of $\M$ by a deterministic quantity:
\begin{equation}
\left\langle G_{ij}(z) \right\rangle = \frac{\delta_{i,j}}{z - \frac{c_i}{\str_{\B}(\ttr_{\M}(z))}}.
\end{equation}
All in all, the global law of the resolvent of the free multiplication $\M = \C^{1/2} \O \B \O^\dag \C^{1/2}$ is given by 
\begin{eqnarray}
z \left\langle G_{ij}(z) \right\rangle & = & \delta_{i,j} Z(z) \left( Z(z) - c_i \right)^{-1}\,,\nonumber \\
Z(z) & \deq & z \str_{\B}(z\stj_{\M}(z) - 1)
\end{eqnarray}
which is exactly the result stated in Eq.\ (\ref{global_law_multiplication})  since the indexes $i,j$ were arbitrary throughout the computations.

\subsection{Information-Plus-Noise matrix}
\label{sec:appendix_info}

The computation of the global law estimate for this model is pretty similar to the sample covariance \cite{bouchaud2003theory}. The noisy measurement matrix is given by
\begin{equation*}
\M = \frac1T (\AAA + \sigma \X)(\AAA + \sigma \X)^{\dag}
\end{equation*}
with $\AAA$ a fixed $N \times T$ matrix such that $T^{-1} \AAA \AAA^{\dag} = \C$. {As we posit that $\X$ is a Gaussian matrix, we can work in the basis where $\C$ is once again diagonal}. Moreover, we can in fact show that interchanging the integral and the average leads to the same result. We hence consider directly the annealed average and abbreviate $\G(z) \equiv \G_{\M}(z)$ in order to lighten the notations. Let us compute the entries of $\G$ using \eqref{eq:replica_trick} with $n=1$ and we find
\begin{align*}
& \left\langle G_{ij}(z) \right\rangle = \int \left(\prod_{k=1}^{N} \dd\eta_k\right) \eta_i \eta_j e^{-\frac{z}{2} \sum_{k=1}^{N} \eta_k^2 } \times \\
& \qquad \qquad\qquad\left\langle e^{\frac{1}{2T} \sum_{k,l=1}^{N} \sum_{t=1}^{T} \sigma_t^2 \eta_k c_k^{1/2} Y_{k,t} Y_{l,t} c_l^{1/2} \eta_l} \right\rangle,
\end{align*}
where we have defined $Y_t := \AAA_t + \sigma \X_t \in \mathbb{R}^{N}$ which is still a Gaussian vector. One can readily compute the average value over the measure of $Y$ to find 
\begin{align*}
& \left\langle \exp\left\{ \frac{1}{2T} \sum_{k,l=1}^{N} \sum_{t=1}^{T} \sigma_t^2 \eta_k c_k^{1/2} Y_{k,t} Y_{l,t} c_l^{1/2} \eta_l \right\} \right\rangle \propto \\
& \qquad\left(1 - \frac{\sigma^2}{T} \eta^{\dag} \eta\right)^{-\frac{T}{2}} \, \exp\left\{\frac{1}{2} \pB{1- \frac{\sigma^2}{T} \eta^{\dag} \eta}^{-1} \sum_{k=1}^{N} c_k \eta_k^2 \right\},
\end{align*}
where we omitted all constants terms and used Sherman-Morrison formula in the exponential term.  Rewriting $p = \sigma^2 T^{-1} \eta^{\dag} \eta$ that we enforced by a Dirac delta function, we can therefore compute the integral over $\{ \eta_k \}$ to find
\begin{equation}
\label{eq:global_law_signalnoise_tmp}
\left\langle G_{ij}(z) \right\rangle = \int \dd p \int \dd\zeta \frac{\delta_{i,j}}{z - q\zeta \sigma^2 - \frac{c_i}{1-p^*}} \exp\left\{- \frac{N}{2} F_0(p, \zeta) \right\},
\end{equation}
and we can once again compute the integral in the large $N$ limit by performing the saddle-point of the following free energy 
\begin{align}
& F_{0}(p, \zeta) \;\deq\; \frac{1-q}{q} \log(1-p) - p \zeta \\
& \qquad\qquad\qquad + \; \frac1N \sum_{k=1}^{N} \log\left[ (z-q\sigma^2\xi)(1-p) - c_k \right]\,.\nonumber
\end{align}
The derivation over $\zeta$ gives the following equation $p^* = q\sigma^2(1-p^*) \stj_{\C}\left( (z - q \zeta \sigma^2)(1-p^*) \right)$, and by taking the normalized trace of the resolvent, we see that we have 
\begin{equation}
 	p^* = q\sigma^2 \stj_{\M}(z).
\end{equation}  
The other derivative leads to 
\begin{equation}
	\zeta^* = \frac{1-q}{q} + z \stj_{\M}(z).	
\end{equation}
Therefore, by plugging $\zeta^{*}$ and $p^*$ into \eqref{eq:global_law_signalnoise_tmp} and then by sending $N \to \infty$ leads to: 
\begin{equation}
\left\langle G_{ij}(z) \right\rangle = \delta_{i,j} \left( (z Z(z) - \sigma^2(1-q)) - Z(z)^{-1} \C \right)^{-1}_{ii} ,
\end{equation}
with
\begin{equation}
Z(z) = 1 - q \sigma^2 \stj_{\M}(z).
\end{equation}
This is exactly the result announced in Eqs. (\ref{global_law_info_noise}) and (\ref{Z_info_noise}) or in \cite{hachem2013bilinear}, showing that considering directly the annealed average is indeed correct in the limit $N \to \infty$.

\section{Abbreviations and Symbols}
\label{app:abbreviations}



\begin{table}[h]
\label{table_example}
\begin{tabular}{l l}

\bfseries \textbf{Symbol} & \bfseries \textbf{Definition}\\
RMT & Random Matrix Theory \\
ESD & Empirical Spectral Density \\
LSD & Limiting Spectral Density \\
RI & Rotational Invariance \\
RIE & Rotational Invariant Estimator \\
SNR & Signal to Noise Ratio\\
$\G_\M(z)$ & Resolvent of $\M$, see Eq.\ \eqref{resolvent} \\
$\rho_\M(z)$ & LSD of $\M$ \\
$\stj_\M(z)$ & Stieltjes transform of $\rho_\M$, see Eq.\ \eqref{app_stieltjes} \\
$\btr_\M(z)$ & Functional inverse of $\stj_\M(z)$, see Eq.\ \eqref{app_blue}\\
$\rtr_\M(z)$ & $\rtr$-transform of $\M$, see Eq.\ \eqref{app_red}\\
$\ttr_\M(z)$ & Moment generating function of $\rho_\M$, see Eq.\ \eqref{app_T}\\
$\str_\M(z)$ & $\str$-transform of $\M$, see Eq.\ \eqref{app_S}\\
\end{tabular}
\end{table}

\end{document}